\documentclass[twocolumn]{aastex62}

\usepackage{graphicx}
\usepackage{amsmath}
\usepackage{natbib}

\usepackage{color}

\shorttitle{Dust Transport and Processing in Disk Winds}
\shortauthors{Giacalone et al.}

\begin{document} 

\title{Dust Transport and Processing in Centrifugally Driven Protoplanetary Disk Winds}
\author[0000-0002-8965-3969]{Steven Giacalone}
\affil{Department of Astronomy, University of California Berkeley, 501 Campbell Hall, Berkeley, CA 94720, USA}
\author[0000-0001-6683-4188]{Seth Teitler}
\affil{National Academic Quiz Tournaments, LLC, 11521 W 69th Street, Shawnee, KS 66203, USA}
\author[0000-0002-6827-1005]{Arieh K\"onigl}
\affil{Department of Astronomy and Astrophysics,The University of Chicago, 5640 S Ellis Ave, Chicago, IL 60637, USA}
\author[0000-0002-3291-6887]{Sebastiaan Krijt}
\altaffiliation{Hubble Fellow.}
\affil{Department of Astronomy and Steward Observatory, The University of Arizona, 933 N Cherry Avenue, Tucson, AZ 85721, USA}
\author[0000-0002-0093-065X]{Fred J. Ciesla}
\affil{Department of the Geophysical Sciences, The University of Chicago, 5734 S Ellis Avenue, Chicago, IL 60637, USA}

\begin{abstract}
There is evidence that protoplanetary disks---including the protosolar one---contain crystalline dust grains on spatial scales where the dust temperature is lower than the threshold value for their formation through thermal annealing of amorphous interstellar silicates. We interpret these observations in terms of an extended, magnetocentrifugally driven disk wind that transports grains from the inner disk---where they are thermally processed by the stellar radiation after being uplifted from the disk surfaces---to the outer disk regions. For any disk radius $r$ there is a maximum grain size $a_\mathrm{max}(r)$ that can be uplifted from that location: grains of size $a$\,$\ll$\,$a_\mathrm{max}$ are carried away by the wind, whereas those with $a$\,$\la$\,$a_\mathrm{max}$ reenter the disk at larger radii. A significant portion of the reentering grains converge to---and subsequently accumulate in---a narrow region just beyond $r_\mathrm{max}(a)$, the maximum radius from which grains of size $a$ can be uplifted. We show that this model can account for the inferred crystallinity fractions in classical T Tauri and Herbig Ae disks and for their indicated near constancy after being established early in the disk evolution. It is also consistent with the reported radial gradients in the mean grain size, crystallinity, and crystal composition. In addition, this model yields the properties of the grains that remain embedded in the outflows from protoplanetary disks and naturally explains the inferred persistence of small grains in the surface layers of these disks.

\end{abstract}


\section{Introduction}
\label{sec:introduction}
Spectroscopic mid-infrared observations have provided evidence for the presence of $\sim$0.1--1\,$\micron$ crystalline silicate grains in the surface layers of protoplanetary disks in objects ranging in mass from brown dwarfs to Herbig Ae/Be stars \citep[e.g.,][]{vanBoekel+05,Apai+05,Bouwman+08,watson+09,Sargent+09}. The inferred mean crystalline mass fractions ($\sim$10--20\%) are evidently established at an early phase of the disk evolution (on timescales $\le$1\,Myr) and change little with age \citep[e.g.,][]{Sicilia-Aguilar+09,Oliveira+11}. To reconcile these observations with the fact that silicate grains are found to be largely amorphous in the interstellar medium, it has been proposed that these particles undergo thermal annealing after being incorporated into the disk \citep[e.g.,][]{Gail98}. This process requires temperatures $\gtrsim 10^3$\;K, but the crystalline grains are detected out to distances where the dust temperature is significantly lower: this has motivated models in which the crystalline grains are processed in the hot, innermost region of the disk and are subsequently transported outward. Although it is conceivable that the grains observed farther out were processed locally in shock waves \citep[e.g.,][]{HarkerDesch02}, there is evidence from interferometric measurements for the presence of a radial gradient in crystallinity between the inner ($\la$\,1\,au) and the outer ($\ga$\,10\,au) disk regions \citep[e.g.,][]{vanBoekel+04,Schegerer+08}, which is consistent with the outward-transport interpretation. Similar inferences about the presence and distribution of silicate crystals have been made in the solar system based on data from short-period comets (which originate at distances $>$\,30\,au). In particular,  a crystallinity fraction of up to $\sim$\,35\% was inferred in the context of the Deep Impact mission to Comet 9P/Tempel 1 \citep[][]{Harker+07}, and a major fraction of the $\gtrsim$\,1\,$\micron$ particles in the material returned by the Stardust mission from Comet 81P/Wild~2 \citep[][]{Brownlee+06,McKeegan+06,Zolensky+06} were determined to be crystalline silicates. In the latter case, the presence of a 20-$\micron$ particle that resembles a meteoritic calcium- and aluminum-rich inclusion (CAI) in one of the samples points to thermal processing that took place when the Sun was still very young.

The proposed radial transport scenarios can be broadly divided into two classes: those involving physical mechanisms that operate inside the disk, and those that involve uplifting dust grains and propelling them to larger radial distances outside the disk. Several mechanisms have been suggested as candidates for internal transport, including large-scale flows associated with angular momentum redistribution in the disk, particle diffusion, density waves in gravitationally unstable systems, and photophoresis (see, e.g., \citealt{Ciesla11} for a summary). While each of these mechanisms could in principle carry particles outward, it remains unclear to what extent the specific conditions that are required for their operation are met in actual disks. For example, models invoking vertical diffusion as well as radial advection by gas flows attributable to an effective viscosity \citep[e.g.,][]{Ciesla07,Ciesla10a,Ciesla10b} often associate these two processes with the presence of turbulence within the disk. However, recent observational \citep[e.g.,][]{Flaherty+15,Flaherty+17,Pinte+16} and theoretical \citep[e.g.,][]{Gressel+15,Bai17} studies have indicated that extended regions in protoplanetary disks may have nonturbulent surface layers.\footnote{The above-cited observational studies place an upper limit of  $\sim$\,$10^{-3}$ on the turbulence parameter $\alpha$ (see Appendix~\ref{sec:AppB}) on spatial scales $\ga$\,30\,au, whereas the referenced theoretical papers infer laminar flows in the inner disk regions (down to $\sim$\,1\,au).} In the external transport scenario, dust grains are uplifted from the disk surface and flung radially out by a gas outflow driven by a magnetic \citep[e.g.,][]{Shu+96} or a radiation-pressure \citep[][]{Vinkovic09} force. The dust particles are coupled to the gas by collisional drag and are naturally sorted according to their size $a$ \citep[e.g.,][]{LiffmanBrown95}: particles whose sizes exceed a critical value $a_{\mathrm{max}}$ (see Equation~\eqref{eq:amax}) remain in the disk because they are too heavy to overcome the stellar tidal gravitational force, those with sizes $a\ll a_{\mathrm{max}}$ are readily uplifted and carried away by the outflow, and those with intermediate sizes leave the disk and start moving outward but eventually succumb to the tidal force (which increases linearly with height above the midplane) and reenter the disk.\footnote{In the radiation-pressure transport scenario presented in \citet{Vinkovic09}, the grains that are uplifted from the surface layers of the inner disk are replenished with new grains through turbulent diffusion from the disk interior: this model is thus also subject to the caveat that the region under consideration may not be turbulent.}
 
The \citet{Shu+96} magnetic outflow model and its subsequent elaborations \citep[e.g.,][]{Shu+01} were based on the X-wind picture \citep[e.g.,][]{Shu+00}, wherein a stellar magnetic field threading the disk launches a centrifugally driven outflow from the vicinity of the outer boundary of the stellar magnetosphere (at a distance $<$\,0.1\,au from the star). This model (which addressed the origin of CAIs and chondrules in the solar system) was critiqued by \citet{Desch+10}, who argued that it is not viable at the proposed location of the base of the wind: among other objections, they pointed out that at that distance the temperature would be too high for solid material to exist. Some of these arguments can also be made against a possible application of this model to the interpretation of crystalline silicates: in particular, the dust (subscript~d) temperature $T_\mathrm{d}$ at the edge of the magnetosphere, both above \citep{Hill+01} and inside \citep{Muzerolle+03} the disk, is expected to lie above the sublimation temperature of silicate grains ($T_\mathrm{sub}$$\sim$1300--1700\,K, where the lower and upper limits of the indicated range correspond, respectively, to olivine and pyroxene compositions; e.g., \citealt{Kobayashi+11}).\footnote{Note in this connection that condensation of evaporated dust is another possible mechanism (besides annealing) for the formation of crystalline silicates \citep[e.g.,][]{Gail04}. However, given the high gas temperatures that are expected in the wind (see Section~\ref{subsec:dynamics}), it is unlikely that any grains would form in this way close enough to the disk surface to end up back in the disk. We therefore do not consider dust recondensation in this work.} Magnetocentrifugal winds could, however, be launched from extended portions of a protoplanetary disk along {\em interstellar} magnetic field lines that are trapped when the natal molecular cloud core undergoes gravitational collapse \citep[e.g.,][]{KoniglPudritz00,KoniglSalmeron11}. Observational evidence for the existence of powerful disk outflows of this type has been accumulating for many years \citep[e.g.,][]{Frank+14,Bjerkeli+16,Hirota+17,Fang+18,Banzatti+19}, and it is now believed that they play a central role in transporting the excess angular momentum of the bulk of the matter that accretes onto the star. Recent theoretical work has indicated that this transport can be effective even if the magnetic pressure remains much smaller than the thermal pressure near the midplane of the disk \citep[e.g.,][]{Turner+14}. Such winds are expected to be strong enough to uplift grains from the disk surface \citep{Safier93}, and there is growing evidence that significant amounts of dust are indeed present in actual protostellar disk outflows (e.g., \citealt{BansKonigl12,Ellerbroek+14}; and references therein).

In this paper we investigate the possibility that the inferred distribution of crystalline silicate grains in protoplanetary disks (including the one that surrounded the young Sun) can be attributed to the action of a magnetocentrifugal wind that is launched beyond the dust sublimation radius $r_\mathrm{sub}$ (the distance below which grains that are exposed to the stellar radiation sublimate). In its basic outline, this model is similar to the one first proposed by \citet{Shu+96}: the drag force exerted by the wind uplifts grains from the disk surface, with particles that originate close enough to the center getting heated to a high temperature through exposure to the stellar radiation and those that are of intermediate size reentering the disk farther out. The principal difference is that, unlike in the X-wind scenario, in this case the dust leaves the disk over an extended region at distances where the temperature is low enough for solid particles to be present. Our aim is to examine whether this model can reproduce the observationally inferred crystallinity properties of these disks for plausible values of the physical parameters. Previous attempts to apply MHD disk outflows to the problem of dust uplifting and thermal processing above the disk were carried out by \citet{SalmeronIreland12} and \citet{Miyake+16}; however, these studies only considered vertical motions and neglected the all-important radial transport of the grains.

We employ a global, semianalytic, MHD solution \citep{Teitler11} to describe the gas dynamics of the disk--wind system and use it as the basis for modeling the dust evolution. We study the transport and thermal processing of dust in the wind zone with the help of a Monte Carlo scheme: the computational methodology for this investigation is outlined in Section~\ref{sec:model}, and the results are presented in Section~\ref{sec:results}. For further insight into this problem, we derive in an appendix a grid-based solution of the transport equations that yields the equilibrium dust distribution for the entire disk--wind domain. The astrophysical implications of this scenario are discussed in Section~\ref{sec:discussion}: besides the radial transport of thermally processed dust to the outer regions of protoplanetary disks, this model makes it possible to characterize the properties of the grains that are carried away by the outflows from these disks and to explain the persistence of small grains in their surface layers. Our findings are summarized in Section~\ref{sec:conclusion}.

\section{Modeling approach}
  \label{sec:model}
The adopted semianalytic disk--wind solution is described in Section~\ref{subsec:disk-wind}. As the dust has negligible impact on the accretion and outflow dynamics, we use the gas density and velocity values obtained from this solution as input data for calculating the trajectories of the grains. The equations governing the grain motions are presented in Section~\ref{subsec:dynamics} under the assumptions that the grain velocities rapidly assume their terminal (time-asymptotic) forms and that collisions between grains can be neglected: the validity of these approximations is considered in Appendix~\ref{sec:AppA}. To solve for the dust distribution in the wind, we use a Monte Carlo scheme (Section~\ref{subsec:MC}) that samples uplifted grains along the disk surface: this scheme is well suited for calculating the properties of the grains that reenter the disk. In the absence of collisions, the evolution of the number density of grains of a given size within the wind can be described by a system of linear, first-order, ordinary differential equations, whose steady-state solution can be obtained numerically. We present this alternative derivation of the grain density distribution in Appendix~\ref{sec:AppB} and demonstrate its good agreement with the Monte Carlo results. Finally, in Section~\ref{subsec:process}, we describe how we treat the thermal processing of uplifted grains.

\subsection{Self-similar Disk--Wind Solution}
\label{subsec:disk-wind}

The solution derived by \citet{Teitler11} represents a steady-state accretion disk that is threaded by a large-scale magnetic field: the field launches a centrifugally driven wind that transports all the excess angular momentum of the accreted matter. The evolution of the magnetic flux is treated self-consistently, with the inward drag exerted by the (mostly neutral) inflowing gas countered by ambipolar diffusion (ion--neutral drag). The model assumes self-similarity in the spherical radial coordinate $R$, which makes it possible to seek a semianalytic solution. A solution of this type was originally obtained by \citet{BlandfordPayne82} for the wind zone: the more general disk--wind model presented by \citet{Teitler11} effectively matches a nonideal-MHD solution for the disk interior to the ideal-MHD wind solution derived by those authors.\footnote{For the comparatively low-luminosity sources that we consider, the effect of the radiation pressure force on the uplifted dust is not strong enough to significantly affect the wind dynamics, so we do not consider this force in deriving the wind solution.} The global disk--wind solution is specified by two dimensionless parameters: $\theta$, the square of the ratio of the isothermal sound speed $c_\mathrm{s}$ to the local Keplerian speed $v_\mathrm{K}$; and $\nu$, the square of the ratio of the midplane Alfv\'en speed to $v_\mathrm{K}$. The particular solution detailed in \citet{Teitler11}, which we adopt in this work, is characterized by $\theta = 1\times 10^{-3}$ and $\nu = 7.744\times 10^{-4}$. The magnitude of quantities such as the gas (subscript~g) velocity $\mathbf{v}_\mathrm{g}$ and mass density $\rho_\mathrm{g}$ in this solution is determined by specifying also the values of two dimensional parameters, which we take to be $M_*$, the mass  of the central star, and $\dot M_\mathrm{g,out}$, the gas mass outflow rate (from both sides of the disk) evaluated between the innermost ($r_{-}$) and outermost ($r_{+}$) disk radii. In particular, at any given location, $v_\mathrm{g}\propto M_*^{1/2}$ and $\rho_\mathrm{g} \propto \dot M_\mathrm{g,out}/M_*^{1/2}$, whereas the magnetic field amplitude $B$ scales as $\dot M_\mathrm{g,out}^{1/2}M_*^{1/4}$. In the applications considered in this paper, we use $M_*$=1\,$M_\sun$, $\dot M_\mathrm{g,out}$=\,$3.50\times10^{-8}\,M_\sun\,\mathrm{yr}^{-1}$, $r_{-}$=\,0.1\,au, and $r_{+}$=\,100\,au. These choices correspond to a midplane magnetic field strength of $\simeq$1.0\,G at 1\,au.

The self-similarity assumption implies that the various physical quantities scale as power laws in $R$ along a spherical ray. Specifically, $v_\mathrm{g}$$\propto$$R^{-1/2}$ for a Keplerian disk, whereas $\rho_\mathrm{g}$ and $B$ scale almost exactly as $R^{-3/2}$ and $R^{-5/4}$, respectively, for the particular solution that we adopt. The normalization of each of these quantities depends on the polar angle of the ray (or, equivalently, on its value of $z/r$, using a cylindrical coordinate system ($r$, $\phi$, $z$)). Since the \citet{BlandfordPayne82} wind solution is incorporated into the solution constructed by \citet{Teitler11}, it is possible to specify any of the two dimensionless parameters that define the wind solution on the basis of the full disk--wind solution: for example, the wind parameters $\lambda$ (normalized specific angular momentum, including the magnetic contribution) and $\kappa$ (normalized mass-to-magnetic flux ratio) take the values 108.4 and $1.27\times10^{-3}$, respectively, in the solution that we employ. In this solution, each decade in radius contributes nearly equally to $\dot M_\mathrm{g,out}$ \citep[see][]{BlandfordPayne82}, but the total mass outflow rate remains small in comparison with the mass inflow rate ($\dot M_\mathrm{g,out}$\,=\,0.10\,$\dot M_\mathrm{g,in}$).

The adopted disk--wind model corresponds to a relatively low value of the midplane (subscript~0) plasma $\beta$ parameter (thermal-to-magnetic pressure ratio), $\beta_0=2\theta/\nu \approx 2.6$, and as such may not be representative of real disks (for which $\beta_0$\,$\gg$\,1). In particular, models of this type are characterized by a rather large inflow speed (of the order of the local speed of sound), which implies unrealistically short disk lifetimes \citep[e.g.,][]{Shu+08}. However, the focus of this paper is on the transport of grains in a disk-driven outflow, for which the semianalytic magnetocentrifugal wind solution provides a convenient representation. We employ the full disk--wind solution to specify the conditions at the base of the outflow and to illustrate the capacity of wind-driving disks to convey dust to the disk surfaces, but the details of the disk interior model are not crucial to our conclusions. We also note that even the particular wind solution that we use may not be representative of real outflows in that models with lower values of $\lambda$ (which are typically slower and denser) have been argued to provide better fits to observations of classical T Tauri (cTT) stellar jets \citep[e.g.,][]{Ferreira+06}. The wind solution that we utilize should nevertheless be adequate for studying the main qualitative aspects of dust transport outside the disk.

The trajectories of the fluid elements that partake in the disk--wind flow have a nested morphology, moving inward roughly in parallel at the outer boundary of the disk and peeling off one after another as the center is approached. One can identify the surface of the disk with the locus of the points where $v_{\mathrm{g},r}$=\,0: above this surface the fluid elements move radially out.\footnote{An alternative identification is with the surface where $v_{\mathrm{g},\phi}$ equals the Keplerian speed $v_\mathrm{K}$=$(GM_*/R)^{1/2}$, which is typically close by \citep[see][]{WardleKonigl93}.} In the self-similar model, the locus of these turning points is a conical surface: it is given by $(z/r)$\,$\approx$\,0.06 in the particular solution that we employ. In this paper we take the base of the wind (subscript~b) to be the surface above which all the dust particles that we consider have $v_{\mathrm{d},r}$$>$\,0. This surface is located just above the $v_{\mathrm{g},r}$=\,0 surface.
 
\subsection{Dust Dynamics}
\label{subsec:dynamics}

Although the underlying model for the gas is self-similar, the velocity and density profiles of the uplifted dust are not. The time-asymptotic velocity components of a dust particle, taken to be a compact sphere of radius $a$ and density $\rho_\mathrm{s} = 3.5\,\mathrm{g\,cm}^{-3}$ \citep[e.g.,][]{WeingartnerDraine01a}, are given by
\begin{equation}
\label{eq:vr}
v_{\mathrm{d},r}(r,z,a)=v_{\mathrm{g},r}(r,z) +\left (\frac{v_{\mathrm{d},\phi}^2}{r} - \frac{GM_*r}{(r^2+z^2)^{3/2}}\right ) t_\mathrm{s}(a)\;,
\end{equation}
\begin{equation}
\label{eq:vphi}
v_{\mathrm{d},\phi}(r,z,a)=v_{\mathrm{g},\phi}(r,z) \;,
\end{equation}
and
\begin{equation}
\label{eq:vz}
v_{\mathrm{d},z}(r,z,a)=v_{\mathrm{g},z}(r,z) - \frac{GM_*z}{(r^2+z^2)^{3/2}}\ t_\mathrm{s}(a)\;,
\end{equation}
where $G$ is the gravitational constant and $t_\mathrm{s} = m_\mathrm{d}v_\mathrm{rel}/F_\mathrm{drag}$ is the particle stopping time due to the drag force $\mathbf{F}_\mathrm{drag}$ induced by its motion (at a speed $v_\mathrm{rel}$) relative to the gas (with $m_\mathrm{d}=(4\pi/3)\rho_\mathrm{s}a^3$ being the grain mass). In the Epstein regime that is relevant to our problem, in which the particle size is smaller than the mean free path of the gas particles, the magnitude of the drag force is given by 
\begin{equation}
\label{eq:Fdrag}
F_\mathrm{drag} \approx \frac{8}{3}\sqrt{2\pi}\left(1+\frac{9\pi}{128}{\cal{M}}^2 \right)^{1/2} \rho_\mathrm{g}c_\mathrm{s}v_\mathrm{rel}a^2\,,
\end{equation}
where ${\cal{M}}\equiv v_\mathrm{rel}/c_\mathrm{s}$ \citep[e.g.,][]{DraineSalpeter79}.\footnote{We include the Mach-number term in Equation~\eqref{eq:Fdrag} since the values of $v_\mathrm{rel}$ in our model are supersonic for sufficiently large values of $z/r$ and $r$, with the (negatively sloped) ${\cal{M}}^2$\,=\,1 surface in the ($z/r,r)$ plane lying closer to the disk surface the larger the particle size $a$.} The sound speed is given by $c_s = (k_\mathrm{B}T_\mathrm{g}/\mu m_\mathrm{H})^{1/2}$, where $k_\mathrm{B}$ is the Boltzmann constant, $m_\mathrm{H}$ is the mass of a hydrogen nucleus, and $\mu$ is the molecular weight; we use $\mu=2.33$ under the assumption of a molecular hydrogen-dominated gas, but our results are not sensitive to the precise value of this parameter. 

We evaluate the gas temperature in the wind using a phenomenological expression of the form
\begin{equation}
\label{eq:Tg}
T_\mathrm{g} = T_1  [(z/r)/(z/r)_\mathrm{b}]^{\gamma} (r/\mathrm{au})^{-\delta}\,,
\end{equation}
and adopt (1)~$T_1$=\,200\,K, $\gamma$\,=\,1, $\delta$\,=\,1/2 as fiducial parameter values based on rough fits to the results of the thermo-chemical model of magnetized protostellar disk winds presented in \citet{Panoglou+12}. That model employed a low-$\lambda$ MHD wind solution corresponding to a relatively dense outflow that provides effective shielding from the protostellar radiation, and it was found that the dominant processes that determined $T_\mathrm{g}$ were the ``generic'' ones associated with weakly ionized, molecular disk winds: ambipolar diffusion heating and adiabatic expansion plus molecular radiation cooling \citep[see][]{Safier93}. It is, however, conceivable that, in the case of more tenuous outflows, X-ray and FUV radiation from the central source could strongly affect the gas temperature. Such effects were previously considered for disks that do not drive outflows \citep[e.g.,][]{Glassgold+04,NomuraMillar05}, and it has been argued that they in fact dominate in certain observed sources \citep[e.g.,][]{Tilling+12,Bruderer+12}. In the case of the young protostar HL~Tau, in which there is evidence for a significant outflow \citep[e.g.,][]{Klaassen+16}, \citet{Kwon+11} inferred a dust temperature profile along the disk ($T_\mathrm{d}(r)$$\approx$\,$600\,(r/\mathrm{au})^{-0.43}$\,K) that is nearly identical to the one predicted for the centrally irradiated surface layer of a cTT disk ($T_\mathrm{d}(r)$$\approx$\,$550\,(r/\mathrm{au})^{-0.40}$\,K; Equation~(11) in \citealt{ChiangGoldreich97}). This estimate likely represents a lower limit on the gas temperature at the top of the disk in this source (see, e.g., the $T_\mathrm{d}$ and $T_\mathrm{g}$ profiles presented in \citealt{Glassgold+04} and \citealt{NomuraMillar05}). Furthermore, as \citet{Panoglou+12} pointed out, $T_\mathrm{g}$ rose much more rapidly with $z/r$ in their magnetocentrifugal wind model than in a hydrostatic disk atmosphere. In the absence of a detailed model of a weakly shielding disk wind, we use the ansatz in Equation~\eqref{eq:Tg} for a qualitative exploration of the possible behavior of the gas temperature in such an outflow. In addition to the fiducial parameter set (1) given above, we consider the following parameter combinations: (2)~$T_1$=\,600\,K, $\gamma$\,=\,1, $\delta$\,=\,1/2---intended to test the effect of a higher temperature at the base of the wind (as indicated in the HL~Tau system); (3)~$T_1$=\,200\,K, $\gamma$\,=\,2, $\delta$\,=\,1/2---intended to test the effect of a steeper vertical temperature gradient; and (4)~$T_1$=\,200\,K, $\gamma$\,=\,1, $\delta$\,=\,2/3---meant to examine the possibility that the radial temperature gradient may also be steeper. For the latter two profiles, we were guided by the results presented in Figure~2 of \citet{NomuraMillar05} for a hydrostatic cTT disk: we inferred $\gamma$\,$\approx$\,2 from a rough fit to the base of the atmosphere in the $r$\,=\,1\,au profile shown in panel~(a\,-1), and $\delta$\,$\approx$\,2/3 from an approximate fit to the $z/r$\,=\,0.1 profile in panel~(a\,-2). The parameter values of the adopted $T_\mathrm{g}$ profiles are summarized in Table~\ref{tab:table1}. At the comparatively low altitudes ($z/r$$<$0.3) attained by the intermediate-size grains that are of interest to us, $T_\mathrm{g}$ likely does not exceed $\sim$$10^3$\,K; however, there is evidence from the analysis of forbidden line emission in protostellar systems that the gas temperature in disk winds can reach much higher values ($\sim$$10^4$\,K) further up \citep[e.g.,][]{Fang+18}. Note, however, that the \citet{BlandfordPayne82} wind model that we employ is ``cold,'' in the sense that the thermal pressure force has a negligible effect on the flow dynamics: this remains true for all the $T_\mathrm{g}$ profiles that we use.
 
 \begin{deluxetable}{lccc}[htb!]
 \tablecaption{Gas Temperature Profiles (Equation~\eqref{eq:Tg})\label{tab:table1}}
 \tablehead{\colhead{Profile} & \colhead{$T_1\,[\mathrm{K}]$} & \colhead{$\gamma$}  & \colhead{$\delta$}}
 \startdata
$1$\tablenotemark{a}  & $200$ & $1$ & $1/2$\\
$2$ & $600$ & $1$ & $1/2$\\
$3$ & $200$ & $2$ & $1/2$\\
$4$ & $200$ & $1$ & $2/3$
 \enddata
 \tablenotetext{a}{Fiducial case.}
 \end{deluxetable}
 
One can estimate the maximum grain size, $a_\mathrm{max}(r)$, that can be uplifted from any given location at the base of the wind by setting the grain velocity component normal to that surface, $v_\mathrm{d,\perp}$$\,=\,$$v_{\mathrm{d},z}$\,-\,$(z/r)_\mathrm{b} v_{\mathrm{d},r}$, equal to zero. Given that $ v_{\mathrm{d},r}$$\approx$\,0 at the base of the wind, the condition $v_\mathrm{d,\perp}$=\,0 reduces to $v_{\mathrm{d},z}$\,$\approx$\,0 at $(z/r)_\mathrm{b}$. Using Equations~\eqref{eq:vz}--\eqref{eq:Tg} in the limit ${\cal M}\ll1$ as well as the self-similarity scalings, which imply $r^2\rho_\mathrm{g}v_{\mathrm{g},z}$\,$\approx$\,$\dot M_\mathrm{g,out}/[4\pi \ln{(r_{+}/r_{-})]}$ at the base of the wind, we obtain
\begin{eqnarray}
\label{eq:amax}
a_\mathrm{max}  \approx & \sqrt{\frac{8}{\pi}} & \frac{\dot M_\mathrm{g,out}}{4\pi \ln{(r_{+}/r_{-})}}\frac{(k_\mathrm{B}T_1/\mu m_\mathrm{H})^{1/2}}{\rho_\mathrm{s}GM_*(z/r)_\mathrm{b}}\left(\frac{r}{\mathrm{au}}\right )^{-{\delta}/{2}} \nonumber \\
=& 0.35&\,\left(\frac{\dot M_\mathrm{g,out}}{10^{-8}\,M_\sun\,\mathrm{yr}^{-1}}\right) \left (\frac{M_*}{M_\sun}\right)^{-1}\left(\frac{T_1}{200\,\mathrm{K}}\right)^{1/2} \\
&\times& \left (\frac{(z/r)_\mathrm{b}}{0.06}\right)^{-1} \left (\frac{\ln{(r_{+}/r_{-})}}{\ln{10^3}}\right )^{-1} \left(\frac{r}{\mathrm{au}}\right )^{-1/4}\micron\, ,
\nonumber 
\end{eqnarray}
where the numerical expression employs the fiducial parameter values. We verified that Equation~\eqref{eq:amax} provides an excellent approximation to the maximum uplifted grain size by evaluating $v_\mathrm{d,\perp}$\,=\,0 numerically: this only changed the numerical coefficient by $\sim$\,6\% (from 0.35 to 0.33). The function $a_\mathrm{max}(r)$ can be inverted to yield $r_\mathrm{max}(a)$, the maximum radius from which grains of size $a$ can be uplifted.

According to the estimate~\eqref{eq:amax}, $a_\mathrm{max} \propto r^{-\delta/2}$. This implies that, if the exponent $\delta$ in Equation~\eqref{eq:Tg} remains close to the fiducial value, the magnitude of $a_\mathrm{max}$ will depend only weakly on radius: for $\delta$\,=\,1/2, it will decrease just by 44\% when $r$ is increased from 1 to 10\,au. Now, as we noted in Section~\ref{sec:introduction}, uplifted grains that reenter the disk have sizes that are close to $a_\mathrm{max}$---much smaller grains are carried away by the outflow. Equation~\eqref{eq:amax} therefore provides a good approximation to the size of reentering grains for a given set of system parameters.

The final equation required for determining the dust dynamics is the conservation relation for the number density $n_\mathrm{d}(a)$ of grains of size $a$ (which in general varies with $r$ and $z$). In a steady state and neglecting grain--grain collisions as well particle diffusion (where the latter omission is consistent with the assumption of a nonturbulent gas flow), it is given by
\begin{equation}
\label{eq:nd}
\mathbf{\nabla \cdot} (n_\mathrm{d}(r,z,a)\mathbf{v}_\mathrm{d}(r,z,a))=0\,.
\end{equation}
We assume that the grains mixed in with the gas that feeds the outflow have sizes in the range 0.005--5\,$\micron$ and a distribution of the form
\begin{equation}
\label{eq:dnda}
\frac{dn_\mathrm{d}}{da}  \propto \left\{
\begin{array}{l} a^{-3.5} \quad\quad 0.005\, \micron \le a <1\, \micron \\ a^{-5.5} \quad\quad\quad 
1\, \micron \le a < 5\, \micron \end{array}\right.
\end{equation}
\citep[see][]{Pollack+85}. We fix the proportionality constant by setting $\rho_\mathrm{d}$\,=\,0.01\,$\rho_\mathrm{g}$ for the matter that enters the wind. Equation~\eqref{eq:nd} is solved explicitly using the grid-based approach described in Appendix~\ref{sec:AppB} but only indirectly in the Monte Carlo scheme presented in Section~\ref{subsec:MC}.

The mass flux of uplifted grains from any radius $r$ along the disk surface can be evaluated from the integral $\int{m_\mathrm{d} (dn_\mathrm{d}/da) v_\mathrm{d,\perp} da}$ using the distribution~\eqref{eq:dnda} and the constraints
\begin{equation}
 \label{eq:cond}
 a<a_\mathrm{max}(r)\quad \mathrm{and} \quad r> r_\mathrm{sub}(a)\, ,
 \end{equation}
 where $r_\mathrm{sub}(a)$ is given by Equation~\eqref{eq:rsub} in Section~\ref{subsec:process}. Equation~\eqref{eq:cond} expresses the requirements (1) that the grain size be smaller than the maximum size that can be uplifted from the chosen location, and (2) that the selected point lie outside the region where grains of this size sublimate. Integrating this flux over the surface area on both sides of the disk yields the total dust mass outflow rate $\dot M_\mathrm{d,out}$.
 
\subsection{Monte Carlo Scheme}
\label{subsec:MC}

Our approach is patterned after the energy-conserving Monte Carlo technique devised by \citet{Lucy99} for calculating radiative equilibria \citep[see also][]{BjorkmanWood01}. In this formulation, photons are sampled from a given source luminosity in the form of equal-energy packets that are strictly monochromatic (so that shorter-wavelength packets contain fewer photons), with the wavelengths chosen so as to correspond to equal selection probabilities based on the source's spectrum. A key feature of the radiative transfer method that underlies this technique is that it enforces energy conservation. In our adaptation, we substitute grains for photons and mass for energy. We replace the wavelengths in the original scheme by the grain sizes, which are sampled using the grain distribution at the base of the wind. The calculation is greatly simplified by the fact that collisional interactions can be neglected for the systems that we consider, which makes the grains analogous to free-streaming photons.

We perform the calculation on a rectangular grid in $(r,z/r)$ space that extends in $r$ from $r_{-}$ to $r_{+}$ and in $z/r$ from $(z/r)_\mathrm{b}$ to 0.5. The radial axis is divided into 100 segments of equal logarithmic size, whereas the vertical axis is divided into 88 segments of equal linear size: the resulting grid is sufficiently fine to ensure that the values of the relevant physical parameters are nearly constant across any given cell. Our goal is to calculate the steady-state grain distribution in the computational domain. To this end, we launch dust and gas packets from the top of the disk and follow their respective motions. Each gas packet has the same mass $M_\mathrm{g}$ (the numerical value of which does not directly affect the calculation), and all the dust packets also have identical masses, fixed at $M_\mathrm{d}$\,=\,0.01\,$M_\mathrm{g}$. A Monte Carlo draw consists of choosing one gas and one dust packet. It is initiated by selecting the launch radius $r_\mathrm{l}$ using the relation
\begin{equation}
\label{eq:xi}
\xi = \frac{\log{(r_\mathrm{l})}-\log{(r_{-})}}{\log{(r_{+})}-\log({r_{-})}}\ ,
\end{equation}
where $\xi$ is a random deviate in the interval $(0,1)$. Upon choosing $r_\mathrm{l}$, we launch a gas packet from that location. We divide the grains into $K$\,=\,200 equal-probability mass bins that are selected at any given radial location by using the expression
\begin{equation}
\label{eq:ak}
\frac{k+0.5}{K} = \frac{\int_{a_{-}}^{a_k}{a^3(dn_\mathrm{d}/da)da}}{\int_{a_{-}}^{a_+}{a^3(dn_\mathrm{d}/da)da}}\ , \quad k=0,1,2...K-1
\end{equation}
(with $a_{-}$=\,0.005\,$\micron$, $a_+$=\,5\,$\micron$, and $dn/da$ given by Equation~\eqref{eq:dnda}) and randomly sampling the integer~$k$ from the interval $(0,K-1)$. This determines the size $a_k$ of the grains, but the corresponding dust packet (which contains $M_\mathrm{d}/m_\mathrm{d}(a_k)$ [$\propto$\,$a_k^{-3}$] particles) is launched only if the conditions given by Equation~\eqref{eq:cond} are satisfied for the given values of $a_k$ and $r_\mathrm{l}$.
 
The velocity field of the gas packets is determined from the adopted magnetocentrifugal wind solution, whereas that of the dust packets (which are characterized by the size of the grains that they contain) is given by Equations~\eqref{eq:vr}--\eqref{eq:vz}. For any given cell through which a packet moves, we approximate the packet's path by a straight line oriented in the direction of the velocity vector at the center of the cell and extending between the packet's entry point into the cell and the point where this line again intersects the cell boundary (which we take to be the exit point from the given cell as well as the entry point into the next cell). Using the length of this line and the magnitude of the adopted velocity vector, we evaluate the transit time of the packet across the cell. For every selected packet, we carry out this procedure for all the cells that the packet traverses between its launch point at the base of the wind and the point where it leaves the grid. We keep track of the cumulative residence times of the gas and dust packets in a 100$\times$88$\times$1 matrix for the former (corresponding to the single gas particle size) and a 100$\times$88$\times$200 matrix for the latter (corresponding to the 200 grain sizes that we consider). Finally, after $N=5\times10^5$ draws, we calculate the mass density ratio $\rho_\mathrm{d}(a_k)/\rho_\mathrm{g}$ for grains of size $a_k$ in any given cell by dividing the corresponding elements of these two matrices (with a further division by 100 implemented to account for the mass difference between dust and gas packets). The total dust-to-gas ratio in the cell is obtained by summing over $k$, whereas the magnitude of the dust density is found from the local dust-to-gas ratio through multiplication by the wind gas density at the center of the cell.

\subsection{Thermal Processing of Uplifted Grains}
\label{subsec:process}

In our model, the crystalline grains that are transported by the wind to the outer regions of the disk are heated to the high temperatures required for annealing only after they enter the wind. Unlike the situation inside the disk, the outflow density is typically too low to assure equilibration of the gas and dust temperatures, so the values of $T_\mathrm{d}$ and $T_\mathrm{g}$ in the wind zone are largely independent of each other. We estimate the temperature of an uplifted grain by focusing on direct irradiation by stellar photons\footnote{Possible additional contributions to the radiative heating are from stellar photons that scatter off other grains in the wind, photons emitted by wind-borne grains, and photons (representing both intrinsic and reprocessed radiation) that originate in the accretion disk.} and assuming that the grain rapidly attains radiative equilibrium with the incident radiation. We approximate the radiation field as a blackbody that is emitted isotropically from a spherical surface of radius $R_*$, and consider two representative cases: A cTT field, characterized by an effective temperature $T_*$\,=\,4000\,K; and a Herbig Ae (HAe) field, for which $T_*$\,=\,10,000\,K. We adopt a single fiducial value for the photospheric radius in both types of star: $R_*$=\,$2\,R_\sun$. With the additional simplification (revisited in Section~\ref{sec:discussion}) that the wind is everywhere optically thin to the stellar photons, the radiative equilibrium condition at a distance $R$ from the star (approximated for this purpose as a point source) can be written as
\begin{equation}
\label{eq:Td}
\int_0^\infty{\kappa_\mathrm{abs}(\nu)B_\nu(T_\mathrm{d})d\nu}=\left(\frac{R_*}{R}\right)^2\int_0^\infty{\kappa_\mathrm{abs}(\nu)B_\nu(T_*)d\nu}\, ,
\end{equation}
where $B_\nu$ is the Planck function at photon frequency $\nu$ and $\kappa_\mathrm{abs}$ is the dust absorption opacity (which in general depends on the grain size $a$). In solving this equation for $T_\mathrm{d}$, we employ the absorption cross sections  tabulated by \citet{WeingartnerDraine01b} for ``smoothed UV astronomical silicate'' spheres.\footnote{Available at\\ https://www.astro.princeton.edu/$\sim$draine/dust/dust.diel.html.}

The sublimation radius can be inferred from Equation~\eqref{eq:Td} by substituting the value of the sublimation temperature for $T_\mathrm{d}$. This gives rise to an expression of the form
\begin{equation}
\label{eq:rsub}
r_\mathrm{sub}(a)=F(T_*,T_\mathrm{sub},a) L_*^{1/2} T_\mathrm{sub}^{-2}\, ,
\end{equation}
where $F$ is a weakly varying function and $L_*$\,=\,$4\pi\sigma T_*^4R_*^2$ is the stellar luminosity (with $\sigma$ being the Stefan-Boltzmann constant). As a function of the grain size, $F$ takes its lowest values for $a\gtrsim$\;1\,$\micron$, a consequence of the comparatively high radiative cooling efficiency of the larger grains (see Table~2 and related discussion in \citealt{BansKonigl12}). In evaluating Equation~\eqref{eq:rsub}, we adopt $T_\mathrm{sub}$=\,1300\,K, appropriate for olivine-type dust. We find that, as $a$ decreases from 1 to 0.1\,$\micron$, $r_\mathrm{sub}$ increases from 0.12\,au to 0.14\,au for cTT systems, and from 0.84\,au to 2.02\,au for HAe disks.

The exact temperature and rate at which amorphous grains anneal depend on their structure and composition. \citet{Ciesla11} examined three representative experimental determinations of this rate as a function of temperature and concluded that, as soon as an amorphous grain is heated to a temperature that exceeds the threshold value $T_\mathrm{ann}$ for the relevant annealing law, it rapidly becomes fully crystalline on account of the exponential dependence of the characteristic annealing time $\tau_\mathrm{ann}$ on $T_\mathrm{d}$,
\begin{equation}
\label{eq:tann}
\tau_\mathrm{ann}^{-1} = \nu_\mathrm{c} e^{-\frac{E_\mathrm{a}}{k_\mathrm{B}T_\mathrm{d}}}
\end{equation}
(where $\nu_\mathrm{c}$ is a characteristic vibrational frequency of the silicate lattice and $E_\mathrm{a}$ is the activation energy). Based on this result, we adopt the approximation that uplifted grains undergo complete annealing if they are heated to a temperature $\ge$\,$T_\mathrm{ann}$, but that they remain amorphous if their temperature remains below this value. This picture implies that there will be an upper limit ($r_\mathrm{ann}$, defined by $T_\mathrm{d}(r_\mathrm{ann})$\,=\,$T_\mathrm{ann}$) on the launch radius of annealed grains. Since $T_\mathrm{ann}$ is generally $<$\,$T_\mathrm{sub}$, $r_\mathrm{ann}$ is $>$\,$r_\mathrm{sub}$ and there will be a range of disk radii for which uplifted grains are annealed (with the specific values of the boundaries $r_\mathrm{sub}$ and~$r_\mathrm{ann}$ of this range depending on the grain size). The value of $T_\mathrm{ann}(a)$ is obtained from the requirement that, when $T_\mathrm{d}$ increases above $T_\mathrm{ann}$, $\tau_\mathrm{ann}$ becomes smaller than the grain transit time across the launch region, which we write in the form 
\begin{equation}
\tau_\mathrm{dyn}=C\,r/v_\mathrm{K}
\label{eq:tdyn}
\end{equation}
(where $C$ is a proportionality constant). Based on our dynamical model calculations, $C$\,$\la$\,10. In view of the comparatively strong radiation field of HAe stars, any given value of $T_\mathrm{d}(a)$ is attained farther out from the star (where the dynamical time is longer) in these systems than in cTT disks, which implies that $T_\mathrm{ann}(a)$ is generally lower in HAe sources. In fact, the incident radiative flux is the dominant factor that determines $T_\mathrm{d}$, so it is an excellent approximation to use a single representative value of $T_\mathrm{ann}$ for each class of sources. In evaluating $\tau_\mathrm{ann}$, we set $\nu_\mathrm{c}$\,=\,$2\times10^{13}\,\mathrm{s}^{-1}$ and $E_\mathrm{a}/k_\mathrm{B}$\,=\,39,100\,K, based on the results obtained by \citet{Fabian+00} for Mg$_2$SiO$_4$ smoke. For the models presented in this paper, we adopt $C$\,=\,1 (the most conservative choice): the $T_\mathrm{ann}$ values inferred for cTT and HAe systems are then 900\,K and 850\,K, respectively.\footnote{For comparison, the corresponding values for $C$\,=\,10 are 850\,K and 800\,K.} For these values of $T_\mathrm{ann}$, $r_\mathrm{ann}$ increases from 0.27\,au to 0.38\,au for cTT systems and from 2.08\,au to 4.37\,au for HAe disks as the grain size decreases from 1 to 0.1\,$\micron$.
 
The Monte Carlo solution enables us to determine the distribution of annealed grains that reenter the disk as well as their origin. For each dust packet that leaves the disk, we record the maximum value of $T_\mathrm{d}$ (which corresponds to the temperature at the launch radius $r_\mathrm{l}$) and classify the associated grains as ``annealed'' (subscript ``ann'') or ``amorphous'' (subscript ``amo'') depending on whether $T_\mathrm{d,max}$ is $\ge$~or~$<$\,$T_\mathrm{ann}$. We keep track of this information until the packet reenters the disk and use it in constructing separate surface distribution functions of the form $\rho_\mathrm{d}(a_k,r)/\rho_\mathrm{g}(r)$ for these two types of grains. We then calculate the radial distribution of the reentering (subscript~r) mass flux, $\sum_k \rho_\mathrm{d}(a_k,r)v_\mathrm{d,\perp}(a_k,r)$ (where all quantities are evaluated at $(z/r)_\mathrm{b}$), for the annealed and the amorphous grains. The corresponding total mass reentry rates, $\dot M_\mathrm{r,ann}$ and $\dot M_\mathrm{r,amo}$, are obtained by integrating these fluxes over the surface area on both sides of the disk.

The representative parameter values adopted in this paper are summarized in Table~\ref{tab:table2}.
\begin{deluxetable}{lc}[htb!]
\tablecaption{Model Parameters\label{tab:table2}}
\tablehead{\colhead{Quantity} & \colhead{Value (cTT/HAe)}}
\startdata
$M_*\,[M_\sun]$ & $1.0$\\
$R_*\,[R_{\sun}]$ & $2.0$\\
$\dot M_\mathrm{g,out}\,[M_{\sun}/\mathrm{yr}]$ & 3.50$\times$$10^{-8}$\\
$T_*\,[\mathrm{K}]$ & $4000$/$10,000$\\
$T_\mathrm{sub}\,[\mathrm{K}]$ & $1300$\\
$r_\mathrm{sub}(a)\,[\mathrm{au}]$\tablenotemark{a}  & 0.12--0.14/0.84--2.02\\
$T_\mathrm{ann}\,[\mathrm{K}]$  & $900$/$850$\\
$r_\mathrm{ann}(a)\,[\mathrm{au}]$\tablenotemark{a}  & 0.27--0.38/2.08--4.37\\
 \enddata
 \tablenotetext{a}{For grain sizes $a$ in the range 0.1--1\,$\micron$; $r_\mathrm{sub}(a)$ and $r_\mathrm{ann}(a)$ are inversely correlated with $a$.}
 \end{deluxetable}

\section{Results}
  \label{sec:results}

\begin{figure*}[htb!]
\includegraphics[width=0.89\textwidth]{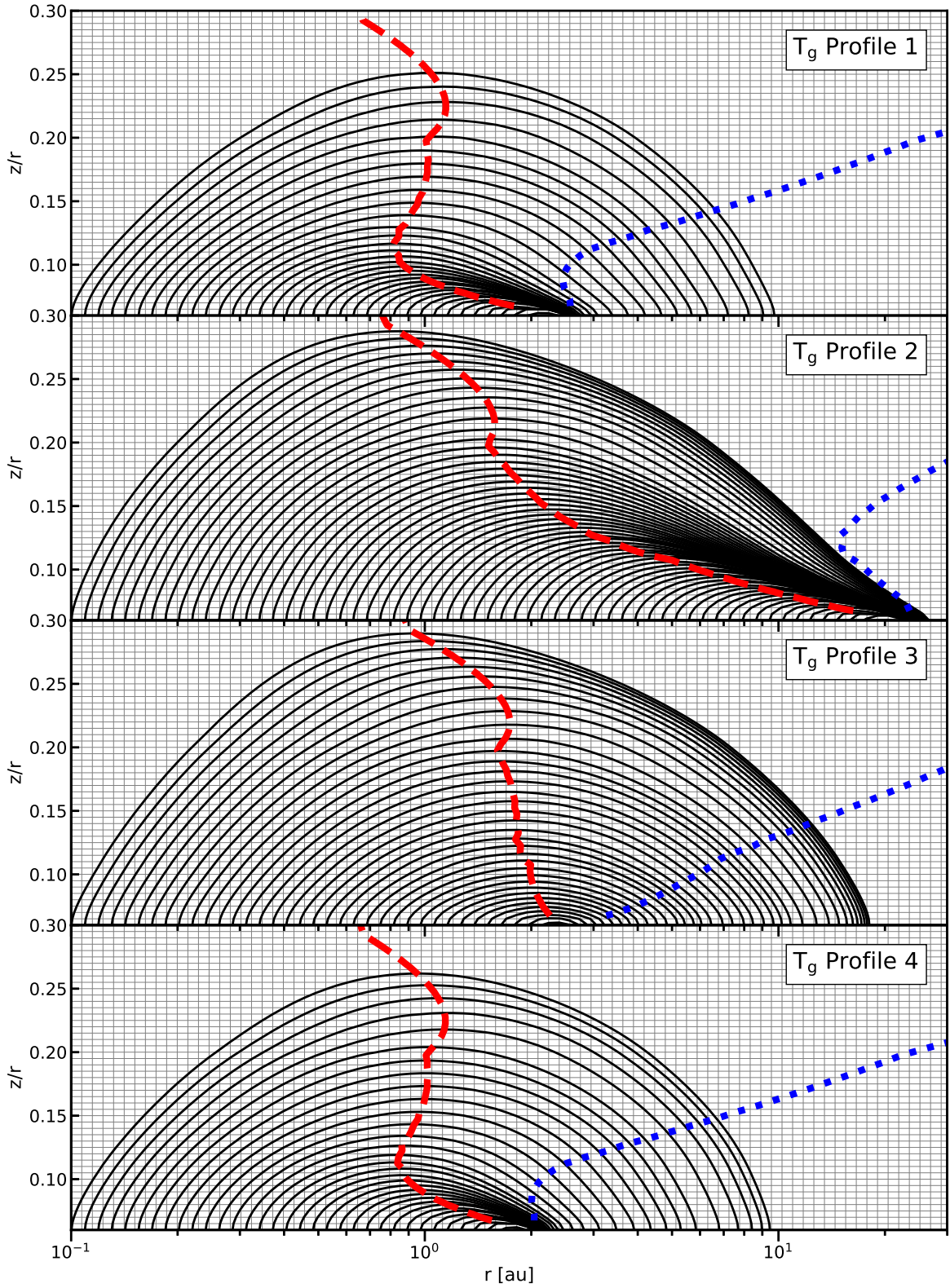}
\caption{Trajectories of uplifted 1\,$\micron$ grains in the $(r,z/r)$ plane for four illustrative profiles of the gas temperature in the wind (see Table~\ref{tab:table1}). The dashed red curve represents the locus of points where the grain velocity component normal to the disk surface vanishes ($v_\mathrm{d,\perp}$\,=\,0), whereas the dotted blue curve represents the locus of points where $v_{\mathrm{d},z}$\,=\,0. The background cells depict the computational grid in the region under consideration.} 
\label{fig:fig1}
\end{figure*}

According to the model outlined in Section~\ref{sec:model}, the presence of $\sim$\,0.1--1\,$\micron$ crystalline silicate grains in the outer regions of protostellar disks around solar-mass stars could in principle be interpreted in terms of the vertical uplifting and radial transport of such grains by hydromagnetic disk winds. To gain insight into this mechanism, we show in Figure~\ref{fig:fig1} the trajectories of uplifted dust particles of size $a$\,=\,1\,$\micron$. For this and the other figures presented in this section, we employ the MHD wind solution described in Section~\ref{subsec:disk-wind} and the nominal parameter values listed in Table~\ref{tab:table2}.  We present results for the four illustrative gas temperature profiles considered in Section~\ref{subsec:dynamics}; however, in this figure we ignore the effect of dust sublimation, so the innermost launch radius is $r_{-}$\,=\,0.1\,au in all cases. The grain trajectories were calculated using Equations~\eqref{eq:vr}--\eqref{eq:vz}, and (in view of the radial self-similarity of the underlying wind model) are plotted in the ($r,z/r$) plane. In each panel, the region to the left (resp., right) of the dashed red curve corresponds to motion away from (resp., toward) the disk surface (defined by $(z/r)_\mathrm{b}$\,$\approx$\,0.06), whereas the dotted blue curve similarly divides the plane into regions where the grains move vertically up or down. The figure demonstrates that all uplifted grains of this size end up back in the disk irrespective of which $T_\mathrm{g}$ profile is employed, but that the detailed structure of their trajectories depends on the choice of parameters in Equation~\eqref{eq:Tg}. The most noticeable difference is in the shape of the trajectories for Profile~2 as compared to those for the three other profiles. In all cases, some fraction of the trajectories are seen to converge to a region---which we term the ``convergence zone''---that lies roughly between the intersection points of the dashed red and dotted blue curves with the base of the wind. For Profile~2, essentially all the reentering grains land within the convergence zone, whereas in the other cases the convergence is less pronounced (particularly for Profile~3). The convergence behavior is exhibited primarily by reentering grains that approach the (tilted) disk surface while still moving upward, whereas the grains that overshoot the convergence zone all move on highly curved trajectories and reenter the disk with $v_{\mathrm{d},z}$\,$<$\,0.

The inner boundary of the convergence zone corresponds to $r_\mathrm{max}(a)$, the outermost radius from which grains of size $a$ can be uplifted. This 
is, in fact, the innermost radius where any grain of this size can reenter the disk since the terminal velocity vectors of these grains are directed away from the disk surface for all radii $r$\,$<$\,$r_\mathrm{max}(a)$. According to Equation~\eqref{eq:amax}, $r_\mathrm{max}(a)$\,$\propto$\,$T_1^{1/\delta}$: this explains why, in comparison with the situation for Profile~1 (the fiducial case), the convergence region is located further out for Profile~2 and further in for Profile~4 (which correspond, respectively, to higher values of $T_1$ and $\delta$; see Table~\ref{tab:table1}). The large (factor of~9) increase in the value of $r_\mathrm{max}(a)$ for Profile~2 in comparison with the fiducial profile is a key reason for why the reentering grains do not overshoot the convergence zone in this case: this region is located far enough from the center that even grains uplifted from the innermost disk region can reach it while still moving upward. The apparent weakening of the convergence behavior for Profile~3 as compared with the fiducial profile can, in turn, be attributed to the more rapid vertical increase of $T_\mathrm{g}$ in this case (a consequence of the higher value of the parameter~$\gamma$). The higher gas temperature leads to a stronger coupling between the dust and the gas, which increases the magnitude of  $v_{\mathrm{d},z}$ above the disk surface: this has the effect of ``unbending'' the trajectories of particles that, in the fiducial case, converge toward $r_\mathrm{max}(a)$.

\begin{figure}[htb!]
\includegraphics[width=0.5\textwidth]{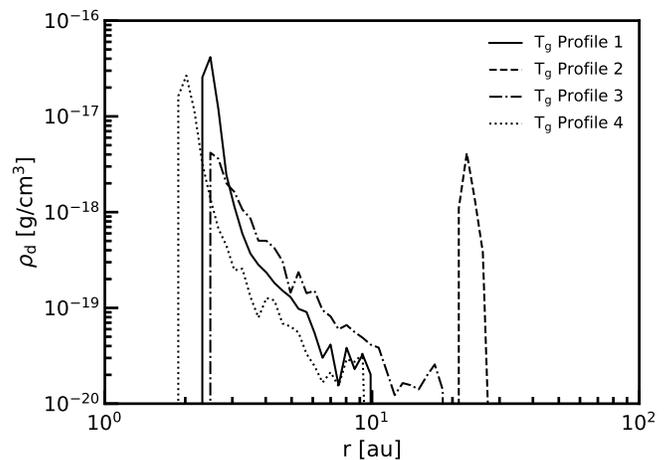}
\caption{Mass density of reentering 1\,$\micron$ grains along the base of the wind for the same four cases shown in Figure~\ref{fig:fig1}.}
\label{fig:fig2}
\end{figure}

A clear manifestation of the convergence behavior of the reentering grains in Figure~\ref{fig:fig1} is provided by the radial distribution of their mass density at the base of the wind, which is shown in Figure~\ref{fig:fig2}. For each of the plotted cases, the density rises sharply at the respective value of $r_\mathrm{max}(a)$ and then exhibits a strong drop within the convergence region; for Profiles~1, 3, and~4, this is followed by a less steeply declining ``shoulder,'' which corresponds to the overshooting grains. As could be expected from the trajectory plots in Figure~\ref{fig:fig1}, the density decreases most rapidly for Profile~2 and most slowly for Profile~3; however, even in the latter case there is a distinct $\sim$\,1 magnitude initial drop in density, which delineates the convergence zone.

\begin{figure*}[htb!]
\includegraphics[width=1.0\textwidth]{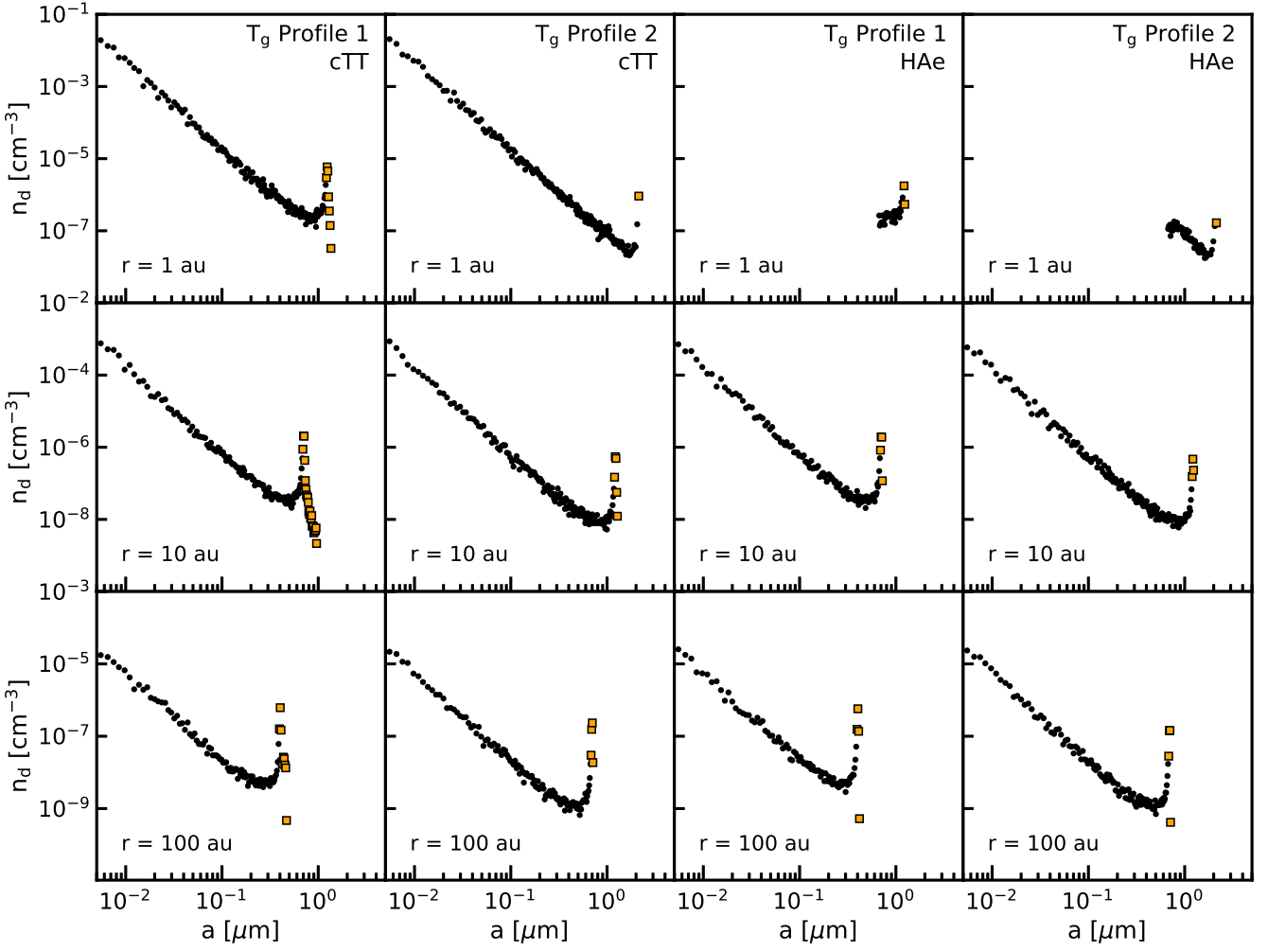}
\caption{Dust number density at the base of the wind for the wind-zone solution, plotted as a function of the grain size $a$ at three different disk radii. The gas temperature profile and the type of radiation field used in determining the grains' sublimation radii are listed for each column of panels. The black circles and orange squares represent outgoing and reentering grains, respectively.} 
\label{fig:fig3}
\end{figure*}

For another perspective on the evolution of the uplifted dust, we plot in Figure~\ref{fig:fig3} the total (comprising outgoing and reentering grains) number density distribution $n_\mathrm{d}(a)$ at three distinct locations ($r$\,=\,1, 10, and~100\,au) along the base of the wind.\footnote{Note that the mass density derived by sampling packets that contain grains of size~$a$ is related to $n_\mathrm{d}$ through a conversion factor that scales as $a(dn_\mathrm{d}/da)$ (see Equation~\eqref{eq:ak}), which we evaluate using Equation~\eqref{eq:dnda}.} Henceforth, a figure panel labeled cTT or HAe indicates that the constraint $r$\,$>$\,$r_\mathrm{sub}$ in Equation~\eqref{eq:cond} is taken into consideration in determining the plotted dust distribution for the respective radiation field (see Table~\ref{tab:table2}). In this figure we show results for $T_\mathrm{g}$ Profiles~1 and~2 and for both cTT and HAe stars. Each of the plotted distributions exhibits the same basic structure for its two components: The outgoing grains (black circles), which occupy the range $a$\,$<$\,$a_\mathrm{max}(r)$; and the reentering grains (orange squares), for which $a$\,$>$\,$a_\mathrm{max}(r)$. The distribution of uplifted grains has the power-law form of the injected dust (Equation~\eqref{eq:dnda}) up to $a$$\,\la\,$$a_\mathrm{max}(r)$, though for sufficiently large radii it turns around and starts to increase with size as $a_\mathrm{max}(r)$ is approached. This rise reflects the fact that, as it gets harder to accelerate the grains, their upward speed decreases, with the effect becoming more pronounced at larger radii because the ratio of the gravitational to the drag-force terms in Equation~\eqref{eq:vz} scales as $r^{\delta/2}$. The reentering dust distribution dominates the uplifted grain distribution when they cross at $a_\mathrm{max}$, but it decreases rather steeply with $a$ for larger grain sizes. 

\begin{figure*}[htb!]
\includegraphics[width=\textwidth]{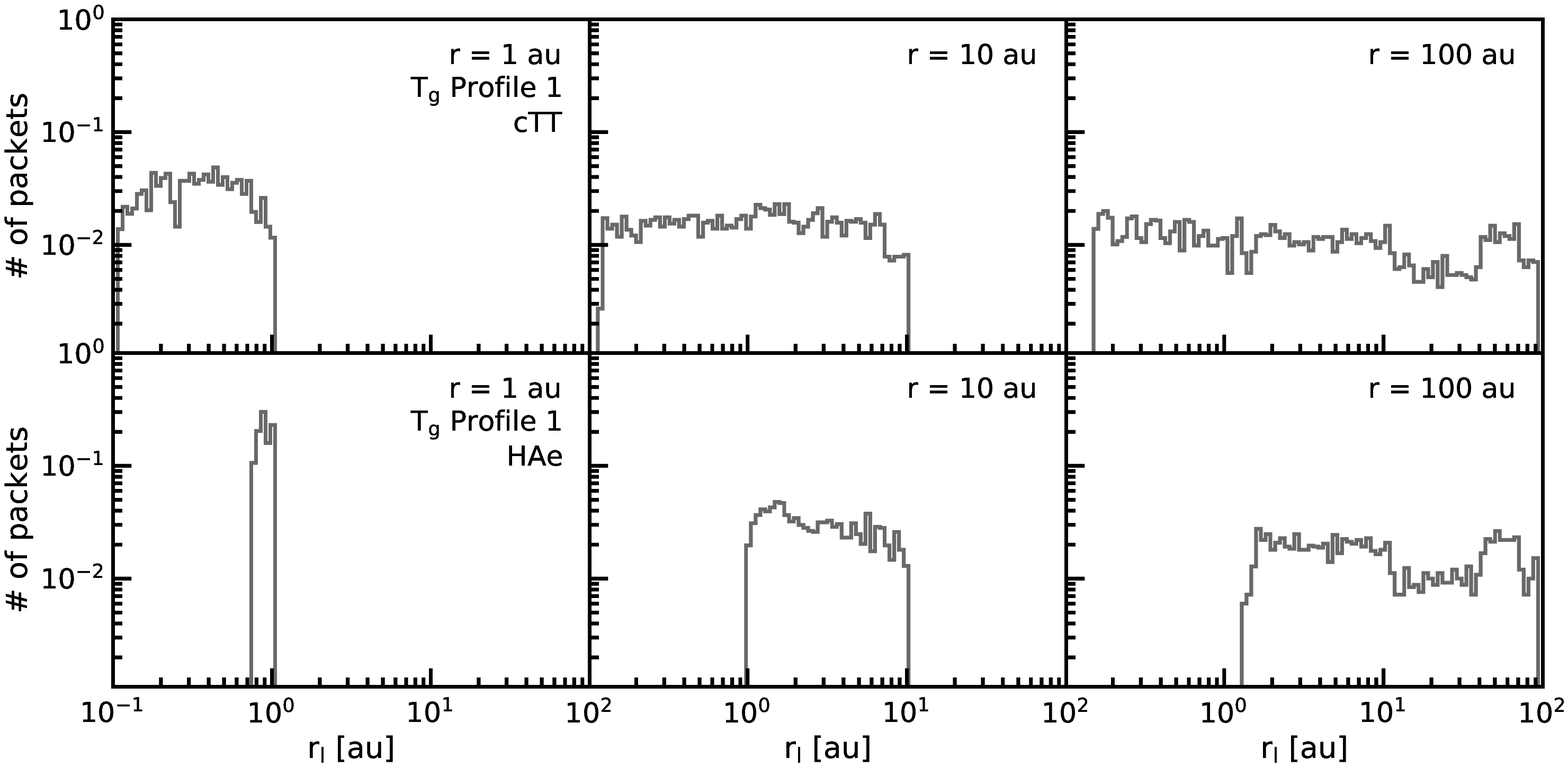}
\caption{Normalized distribution of the reentering dust mass (integrated over al grain sizes) as a function of $\log_{10}{r_\mathrm{l}}$ (the logarithm of the launch radius) in the fiducial model for the same three disk radii considered in Figure~\ref{fig:fig3}. The top and bottom rows correspond, respectively, to the cTT and HAe radiation fields.}
\label{fig:fig4}
\end{figure*}

The different panels of Figure~\ref{fig:fig3} show that the dust distribution at the base of the wind can vary in its detailed shape even as its basic structure remains the same. For any given $T_\mathrm{g}$ profile and radiation field, the upper cutoff of the distribution decreases with radius---this is due to the fact that $a_\mathrm{max}$ is a decreasing function of $r$ (see Equation~\eqref{eq:amax}). The higher value of the cutoff in the case of Profile~2 (as compared with Profile~1) can be similarly attributed to the temperature dependence of this quantity ($a_\mathrm{max}$\,$\propto$\,$T_\mathrm{g}^{1/2}$).\footnote{The density distributions for Profiles~3 and~4 also differ from the fiducial case in the details of their shapes, and these differences, too, can be readily understood from the properties of the grain trajectories. However, as these variations are fairly minor, we do not display the corresponding plots here.} The cTT and HAe distributions differ only in the inner disk region where sublimation effects play a role. In particular, at $r$\,=\,1\,au, the uplifted grain distribution extends all the way down to $a_{-}$=\,0.005\,$\micron$ for the cTT field but only down to $\sim$\,0.7\,$\micron$ when exposed to the more intense HAe radiation. In the latter case, the value of the lower cutoff is consistent with the expectation that irradiated $a\gtrsim$\;1\,$\micron$ grains attain lower temperatures than smaller particles on account of their higher radiative efficiencies (see Section~\ref{subsec:process}).

One can directly infer the mass distribution of reentering grains by tallying the dust packets that return to the disk at any given radius and utilizing the fact that, in our Monte Carlo scheme, each packet has the same mass. Figure~\ref{fig:fig4} shows the (normalized) distributions obtained in this way as a function of the launch radius $r_\mathrm{l}$ for the same three radii employed in Figure~\ref{fig:fig3}. It is seen that the mass distribution is nearly flat in $\log_{10}{r_\mathrm{l}}$. This can be understood by noting that the reentering dust at radius $r$ is dominated by particles with $a$\,$\ga$\,$a_\mathrm{max}(r)$ that arrive from the entire region between $\mathrm{max}\{r_{-},r_\mathrm{sub}\}$ and $r$ (see Figures~\ref{fig:fig1} and~\ref{fig:fig3}) and by recalling that the self-similar wind model corresponds to a nearly equal contribution of mass discharge from each decade in radius (see Section~\ref{subsec:disk-wind}). This is a robust property that is not sensitive to the details of the $T_\mathrm{g}$ distribution, so we only exhibit the fiducial case. However, the amount of mass that reenters the disk is different for cTT and HAe systems on account of the increase in the magnitude of  $r_\mathrm{sub}$  (which, for the adopted parameter values, represents the innermost launch radius in both systems) on going from cTT stars to HAe stars (see Table~\ref{tab:table2}).

\begin{figure*}[htb!]
\includegraphics[width=1.0\textwidth]{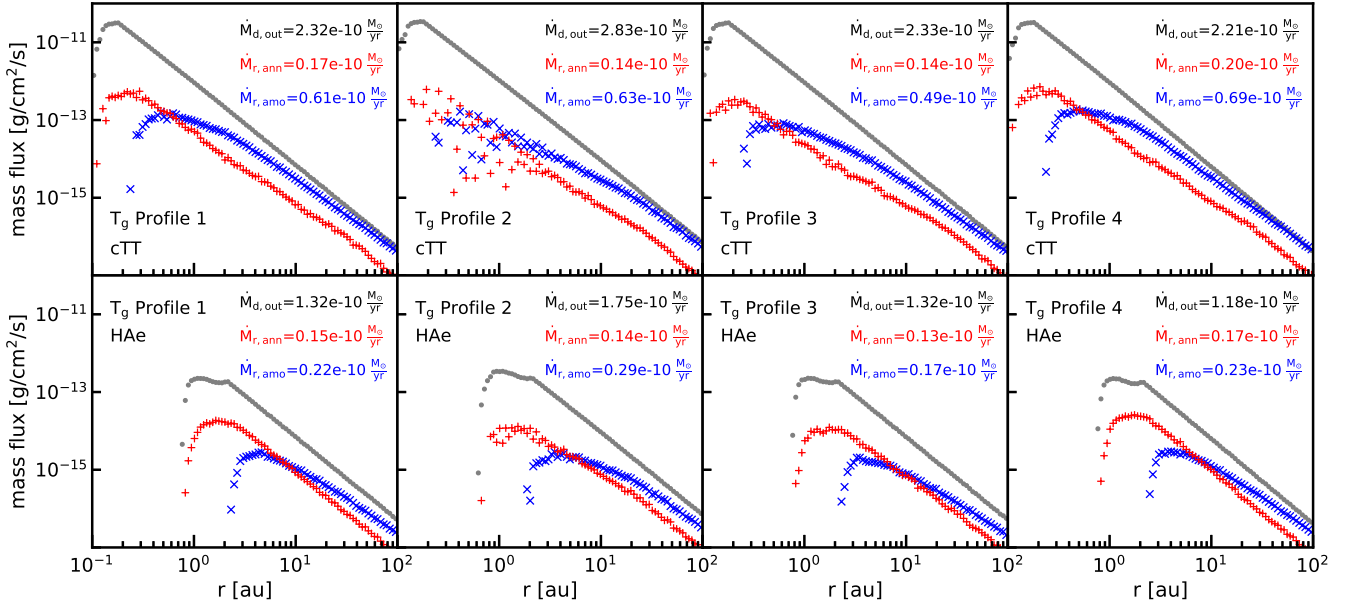}
\caption{Radial distributions of the dust mass fluxes along the base of the wind. The black circles represent \emph{uplifted} grains whereas the red plus signs and blue crosses correspond, respectively, to annealed ($T_\mathrm{d,max}$\,$\ge$\,$T_\mathrm{ann}$) and amorphous ($T_\mathrm{d,max}$\,$<$\,$T_\mathrm{ann}$) \emph{reentering} grains. Each panel lists the relevant $T_\mathrm{g}$ profile and radiation field (cTT or HAe) as well as the values of the corresponding mass flow rates across the two disk surfaces. The values of $T_\mathrm{ann}$ are, respectively, 900\,K and 850\,K for cTT and HAe systems.}
\label{fig:fig5}
\end{figure*}

Figure~\ref{fig:fig5}  presents the distribution of the mass fluxes of uplifted and reentering grains along the base of the wind. In this instance we show the results for all of our adopted wind-zone $T_\mathrm{g}$ profiles and again distinguish between cTT and HAe systems. In each case we integrate the different fluxes over the two disk surfaces and list the resultant mass flow rates in the corresponding panel. This figure corroborates our previous inferences: for each class of objects, the results are not sensitive to the choice of the $T_\mathrm{g}$ profile, but the dust outflow and reentry rates are distinctly lower for the HAe systems. For cTT systems, we find that the ratio of the dust mass outflow rate to $\dot M_\mathrm{g,out}$ is $\sim$\,6--8$\times 10^{-3}$, and that $\sim$\,30--40\% of the uplifted grain mass reenters the disk. The corresponding fractions for HAe disks are $\sim$\,3--5$\times 10^{-3}$ and $\sim$\,20--30\%, respectively. Note that the mass flux of uplifted dust scales as $r^{-2}$ over most of the sampled range, as expected from the self-similarity scaling of the ``carrier'' wind.
 
\begin{figure*}[htb!]
\includegraphics[width=1.0\textwidth]{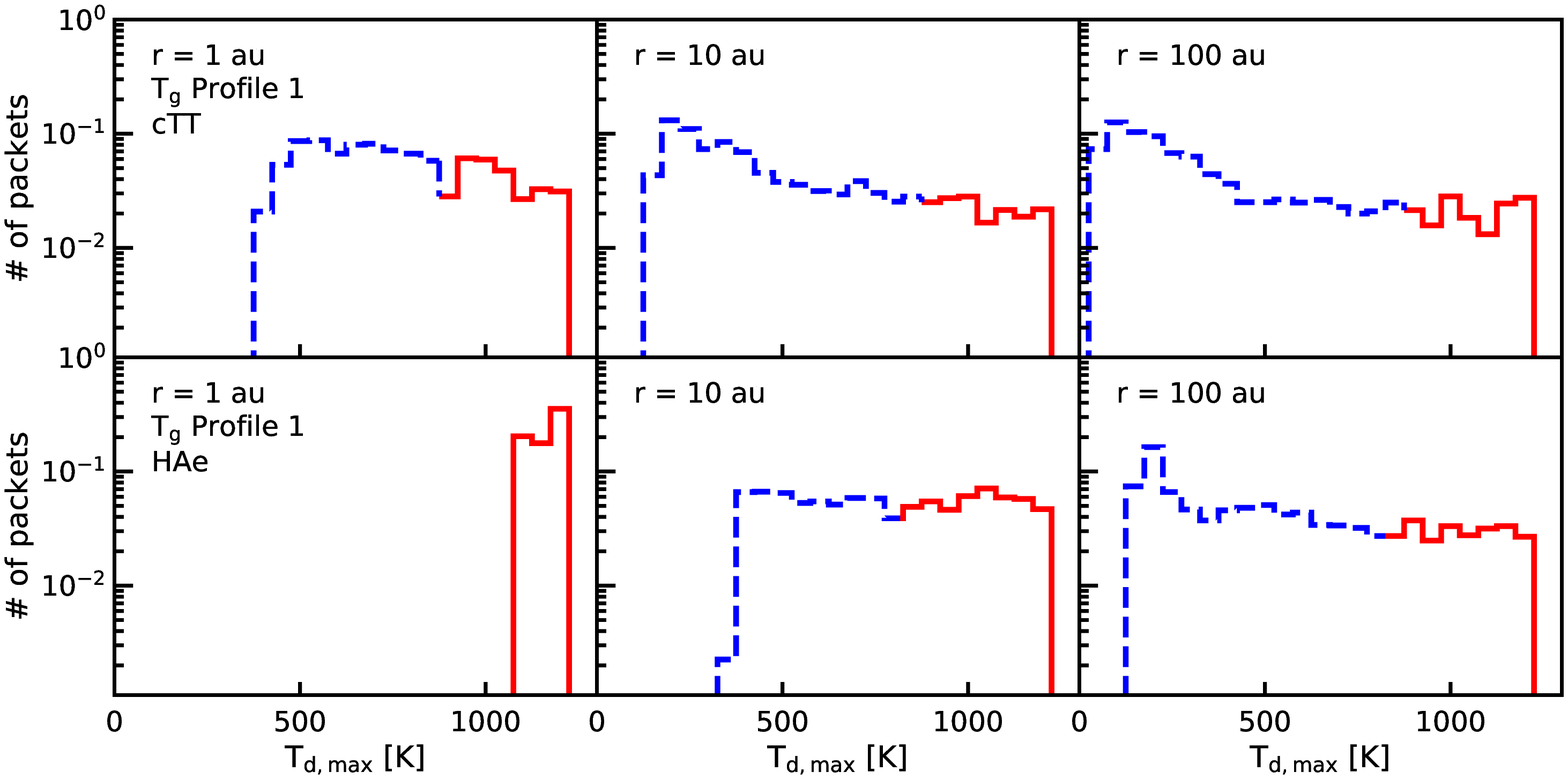}
\caption{Normalized distribution of the reentering dust mass as a function of $T_\mathrm{d,max}$ (the maximum temperature to which grains are heated after they are uplifted from the disk surface) for the same model parameters and disk radii used in Figure~\ref{fig:fig4}. The solid (red) and dashed (blue) portions of each histogram are separated at $T_\mathrm{d,max}$\,=\,$T_\mathrm{ann}$ and correspond to annealed and amorphous grains, respectively.}
\label{fig:fig6}
\end{figure*}

We now consider the model predictions for the transport of annealed grains. In analogy with Figure~\ref{fig:fig4}, one can plot the mass distribution of reentering dust at any given radius as a function of $T_\mathrm{d,max}$: the results for the same three radii considered in that figure are shown in Figure~\ref{fig:fig6}. The overall flattish appearance of these distributions and the progressive decrease of their low-end truncation values with increasing radius essentially mimic the behavior of the mass distributions in Figure~\ref{fig:fig4}.\footnote{Note that $T_\mathrm{d,max}$ in Figure~\ref{fig:fig6} is plotted on a linear scale whereas the $r_\mathrm{l}$ scale in Figure~\ref{fig:fig4} is logarithmic. Note also that, even though $T_\mathrm{d,max}$ is equal to $T_\mathrm{d}(r_\mathrm{l})$ for each individual grain, the mapping between $T_\mathrm{d,max}$ and the launch radius $r_\mathrm{l}$ is not one to one for the full collection. As we already remarked in Section~\ref{subsec:process} in connection with Equation~\eqref{eq:rsub}, the equilibrium temperature of an irradiated grain depends in part on its size: this implies that grains of different sizes that are launched from different radii may arrive at a given location along the base of the wind with the same value of $T_\mathrm{d,max}$.} The histograms in Figure~\ref{fig:fig6} are divided into ``annealed''  (solid red) and ``amorphous'' (dashed blue) portions, which correspond to  $T_\mathrm{d,max}/T_\mathrm{ann}$\,$\ge$\,1 and $<$\,1, respectively.\footnote{Recall from Section~\ref{subsec:process} that $T_\mathrm{ann}(a)$, which is determined from the condition $\tau_\mathrm{ann}(a)$\,=\,$\tau_\mathrm{dyn}$, has different values in cTT and HAe systems.} The contribution of annealed grains dominates in the inner regions of the disk, where most of the arriving grains have $T_\mathrm{d,max}$\,$\ge$\;$T_\mathrm{ann}$, whereas amorphous grains predominate farther out. The transition between these two regimes occurs at $\sim$\,0.5\,au for cTT disks and $\sim$\,5\,au for HAe systems (see Figure~\ref{fig:fig5}), the difference reflecting the stronger dust heating in the more luminous sources. The observational manifestation of this behavior would be a crystallinity gradient with a characteristic spatial scale on the order of the transition radius. It is seen that such gradients arise naturally in this model. 

For the models presented in Figure~\ref{fig:fig5}, the overall crystalline grain fraction in the dust that reenters the disk at $r$\,$\le$\,$10^2$\,au lies in the range $\sim$18--22\% for cTT systems and $\sim$33--43\% for HAe disks.\footnote{The inferred crystallinities are higher when the parameter $C$ in Equation~\eqref{eq:tdyn} exceeds our adopted minimum value ($C$\,=\,1). For example, when $C$\,=\,10, the corresponding ranges are, respectively, $\sim$21--27\% and $\sim$37--48\% for cTT and HAe systems.} The stronger radiation field in HAe systems gives rise to two competing effects: On the one hand, the higher values of $r_\mathrm{sub}(a)$ (Equation~\eqref{eq:rsub}) reduce the total amount of dust that is uplifted from the disk; on the other hand, the higher values of $r_\mathrm{ann}(a)$ (see Section~\ref{subsec:process}) increase the range of disk radii where uplifted grains become annealed. The first effect results in lower values of $\dot M_\mathrm{d,out}$ and, correspondingly, of the total mass reentry rate ($\dot M_\mathrm{r,ann}+\dot M_\mathrm{r,amo}$); the second effect leads to higher values of the ratio $\dot M_\mathrm{r,ann}/\dot M_\mathrm{r,amo}$ and hence of the crystallinity fraction for the reentering dust. The overall conclusion is that, for similar disk extents and mass outflow rates, HAe systems would have distinctly lower total dust reentry rates than their cTT counterparts but only moderately smaller mass reentry rates of annealed grains. The two competing effects are also manifested at the local level: for example, while the mass flux of reentering amorphous grains at $r$\,=\,1\,au in the cTT case is larger than the flux of annealed grains that reenter the disk at that radius in HAe systems, in the latter case the reentering dust is 100\% annealed.

\begin{figure*}[htb!]
\includegraphics[width=1.0\textwidth]{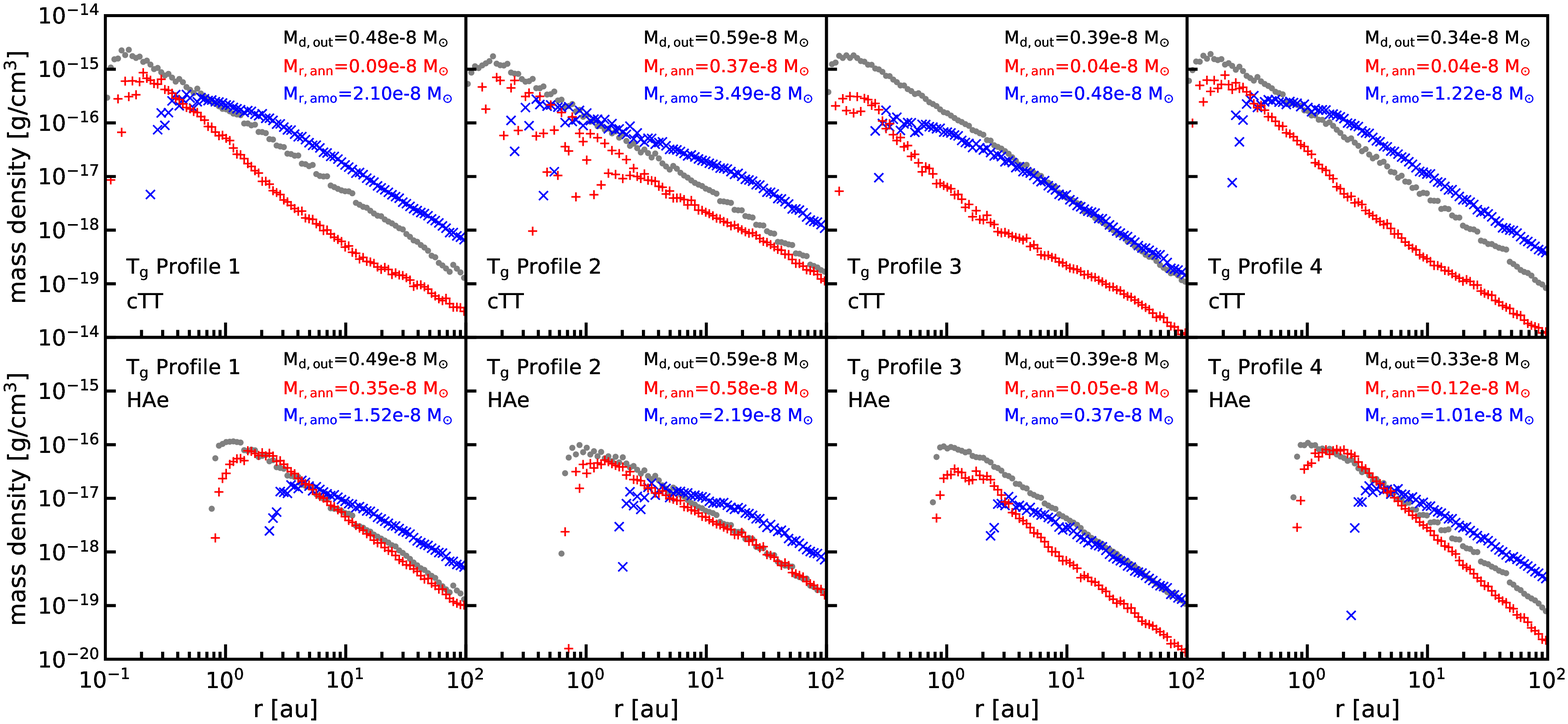}
\caption{Radial distributions of the dust mass densities along the base of the wind. The model outflows presented in this figure as well as the different symbols and colors are the same as those shown in Figure~\ref{fig:fig5}. Besides the relevant $T_\mathrm{g}$ profile and radiation field, each panel also lists the total mass contained in the bottom layer of grid cells on one side of the disk for the three specified dust flow components.}
\label{fig:fig7}
\end{figure*}

The crystallinity fractions in protoplanetary disks are inferred from spectroscopic observations of the dust distribution in the disk surface layer. For a more direct comparison with these observations, we plot in Figure~\ref{fig:fig7} the radial distribution of the mass density of the various dust components at the base of the wind for the four $T_\mathrm{g}$ profiles and two radiation fields employed in this study. The total mass of each of these components within the bottom layer of cells in the computational grid is listed in each panel as a proxy for the dust mass probed by observations.\footnote{In our model, the mass ($M_\mathrm{d,out}$) of the outgoing dust component is invariably higher for cTT systems than for HAe ones because of the smaller range of disk radii from which the outflow originates in the latter systems: the small deviation from this behavior in the case of Profile~1 in Figure~\ref{fig:fig7} is due to undersampling in the Monte Carlo calculation.} The basic appearance of the $\rho_\mathrm{d}$ curves in Figure~\ref{fig:fig7} is similar to that of the $\rho_\mathrm{d} v_{\mathrm{d},\perp}$ curves in Figure~\ref{fig:fig5}. In particular, the outgoing dust component follows the self-similarity scaling of $\rho_\mathrm{g}(r)$ ($\propto$\,$r^{-3/2}$), and the reentering grains exhibit a crossover between the annealed and amorphous components at the same distinct radial locations for cTT and HAe systems as the corresponding curves in Figure~\ref{fig:fig5}. However, while the outgoing dust component dominates the flow of mass through the top of the disk, its contribution to the mass contained at any given time at the base of the outflow is subordinate to that of the reentering grains and constitutes only $\sim$13--43\% (resp., $\sim$18--48\%) of the total dust mass in the layer for cTT (resp., HAe) systems. In view of our previous findings, it it not surprising that the largest value of the reentering dust mass corresponds to Profile~2 and the smallest value to Profile~3. Another difference from the results presented in Figure~\ref{fig:fig5} is in the inferred crystallinity fraction for the reentering dust, which is measurably lower when evaluated on the basis of mass: in this case it lies in the range $\sim$3--10\% for cTT systems and $\sim$11--21\% for HAe disks. The difference from the mass flux-based estimates can be understood from the fact that the magnitude of $v_\mathrm{d,\perp}$ for reentering grains of size $a$ is lower the smaller the value of $a$,\footnote{This can be seen by using Equations~\eqref{eq:vr}--\eqref{eq:vz} to write $v_\mathrm{d,\perp}$ in the form $v_\mathrm{d,\perp}$\,=\,$v_{\mathrm{g},z} - (z/r) v_{\mathrm{g},r} - (z/r)(v_{\mathrm{g},\phi}^2/r) t_\mathrm{s}(a)$, and by noting that $t_\mathrm{s}(a)$\,$\propto$\,$a$ and that $v_\mathrm{d,\perp}$\,$<$\,0 for reentering grains.} and that reentering amorphous grains have on average smaller sizes than their annealed counterparts because they are uplifted from larger radial distances in the disk. Amorphous grains therefore have higher densities than annealed ones for a given mass flux of reentering dust: this is equivalent to the statement that the proportion of annealed grains is higher in the mass flux of reentering grains than in these grains' local mass density. 

\section{Discussion}
  \label{sec:discussion}

Using a magnetocentrifugal wind model, we demonstrated that an extended disk wind can account for many of the observationally inferred properties of annealed grains in protostellar disks, including their characteristic sizes ($\sim$\,0.1--1\,$\micron$) and broad spatial dispersion (to distances $\ga$\;10\,au), as well as the indicated radial crystallinity gradient. Under the assumption that the grains are heated by the stellar radiation to which they become exposed after they rise above the disk surface, this model yields crystallinity fractions for the reentering grains that range from 18\% to 43\% in the examples that we presented (though those values would be lower if opacity effects in the wind reduce the incident radiative flux). To estimate the observed crystallinity fraction, we consider the contribution of annealed grains to the total mass of dust at the base of the wind, which we evaluate using the bottom layer of cells in our computational grid. We get a lower bound---$M_\mathrm{r,ann}/(M_\mathrm{d,out}+ M_\mathrm{r,amo}+M_\mathrm{r,ann})$---on this quantity by taking the outgoing grains to be fully amorphous and neglecting the possible contribution of reentering grains that accumulate below the base of the wind (see below). Using the numerical values listed in the different panels of Figure~\ref{fig:fig7}, this yields lower limits of $\sim$3--8\% for cTT systems and $\sim$7--17\% for HAe disks, which can be compared with the observationally inferred range ($\sim$10--20\%).\footnote{If we chose $C$\,=\,10 in Equation~\eqref{eq:tdyn} instead of $C$\,=\,1 (so that the inferred values of $T_\mathrm{ann}$ for cTT and HAe systems were 850\,K and 800\,K, respectively, rather than 900\,K and 850\,K), the corresponding ranges would be $\sim$3--10\% and $\sim$7--19\%.}

A key feature of wind-transported dust that was identified in our calculations is the convergence of a significant fraction of the reentering grains to a narrow zone at the base of the wind. The two curves in Figure~\ref{fig:fig1} whose intersection with the base defines this zone---$v_\mathrm{d,\perp}$\,=\,0 and $v_\mathrm{d,z}$\,=\,0---themselves intersect at a slightly lower value of $z/r$. The latter intersection point represents a stagnation point of the flow ($v_\mathrm{d}$\,=\,0) for grains of the given size. This suggests that reentering grains whose trajectories pass within their corresponding convergence zones would tend to accumulate just below the base of the outflow. To confirm this inference, we derived the equilibrium dust distribution for the entire disk--wind domain using the grid-based solution method described in Appendix~\ref{sec:AppB}. In this approach, the dust is followed from the outer radial boundary of the disk, where it is envisioned to be part of the interstellar matter that feeds the accretion flow (see Section~\ref{subsec:full}). This solution indeed exhibits high-amplitude density spikes that are centered at the locations of the aforementioned stagnation points, and we verified that the mass flow into the cells where the density peaks is dominated by reentering grains that arrive from the base of the wind. The density spikes represent the time-asymptotic limit of the particle concentration process and thus have unphysically high magnitudes, which in this solution are limited only by numerical diffusivity---in fact, the spikes are characterized by $\rho_\mathrm{d}/\rho_\mathrm{g}$\,$\gg$\,1, which violates the model assumption of gas-dominated dynamics. In real disks, grain-grain collisions and particle diffusion can be expected to intervene before such high densities are attained. As discussed in Appendix~\ref{sec:AppB}, the incorporation of collisional effects into the model results in a large system of nonlinear equations that is prohibitively difficult to solve. However, it is possible to extend the model to include the effect of diffusion, which we have done by including a phenomenological diffusivity (characterized by the commonly employed gas diffusivity parameter $\alpha$) that could represent an underlying gas turbulence (see Section~\ref{subsec:diffusion}). Using this generalized solution, we found that $\rho_\mathrm{d}/\rho_\mathrm{g}$ drops below~1 everywhere in the disk only after $\alpha$ comes to exceed $\sim$\,$1\times 10^{-3}$ (with the transition value remaining nearly constant with $\dot M_\mathrm{g,in}$). While this can be regarded as a moderate level of turbulence, it may still be too high to be consistent with the observational and theoretical indications that large portions of the surface layers of protoplanetary disks are nonturbulent (see Section~\ref{sec:introduction}). Note, however, that grain-grain collisions should be important at the locations where dust reenters the disk (see Section~\ref{subsec:collisions}), so it is likely that collisional effects (particularly coagulations, in view of the low relative grain velocities near the stagnation points) would act to remove dust particles from the convergence zones and thereby limit the growth of the density spikes. The accumulation of reentering grains in their respective convergence zones would have the effect of increasing the observed crystallinity fraction near the base of the wind by enhancing the contribution of this grain population (which is characterized by comparatively high crystallinities) relative to that of the outgoing grains.\footnote{The crystallinity fraction would be further enhanced if---contrary to our assumption that the uplifted grains are fully amorphous---some of the processed grains that reenter the disk are carried back to the disk surface and become part of the outgoing dust component.}

According to Equation~\eqref{eq:amax}, the maximum size of grains that can be uplifted from a given disk radius scales as $\dot M_\mathrm{g,out}/M_*$: it therefore decreases with time as the central mass increases and the mass outflow rate (which tracks $\dot M_\mathrm{g,in}$) goes down. Our model thus implies a direct correlation between the dominant size of reentering grains at a given radius (or the radius where most grains of a given size reenter the disk) and $t_\mathrm{age}$ (the system's age). In particular, grains that reach the outer disk regions at early times could have sizes $\gg$\,1\,$\micron$, which, among other implications, may be relevant to the origin of the 20\,$\micron$ CAI-like particle discovered in one of the samples returned from Comet 81P/Wild~2 (see Section~\ref{sec:introduction}).\footnote{Using a similar argument, \citet{Wong+16} proposed that the inferred presence of millimeter-size grains in protostellar envelopes could be attributed to dust transport by disk winds during an early phase of the protostellar evolution (characterized by $\dot M_\mathrm{g,out}$$\sim$\,$10^{-6}\,M_{\sun}\,\mathrm{yr}^{-1}$ and $M_*$$\sim$\,0.1\,$M_\sun$).} The mass accretion and outflow rates continue to evolve even after the bulk of the protostellar mass has been assembled---in particular, $\dot M_\mathrm{g,in}$ is inferred to decrease with $t_\mathrm{age}$ as $10^{-7}(t_\mathrm{age}/10^6\,\mathrm{yr})^{-1.2}$\,$M_\sun$\,$\mathrm{yr}^{-1}$ for $M_*$ in the range 0.4--1.2\,$M_\sun$ \citep{Caratti+12}. Assuming $\dot M_\mathrm{g,out}$\,$\approx$\,0.1\,$\dot M_\mathrm{g,in}$, it then follows from Equation~\eqref{eq:amax} that the characteristic size of annealed grains varies as $\simeq$\,0.4\,$(M_*/M_\sun)^{-1}(T_1/200\,\mathrm{K})^{1/2}(t_\mathrm{age}/10^6\,\mathrm{yr})^{-1.2}$\,$\micron$: the time evolution of $a_\mathrm{max}$ may contribute to the extent of the observed size range of annealed grains.

Over most of their evolution, protostellar disks remain in a quiescent state wherein $\dot M_\mathrm{g,in}$ and $\dot M_\mathrm{g,out}$ decrease monotonically with $t_\mathrm{age}$, as described above. However, the disks also experience episodes of high accretion and outflow rates, the strongest of which correspond to FU Orionis (FUOR) outbursts \citep[e.g.,][]{HartmannKenyon96}. Mid-infrared observations of FUOR disks \citep{Quanz+06,Green+06} have revealed the presence of amorphous silicate grains, but not of crystalline ones. Though the nature of FUOR outflows is still being debated, it has been argued \citep[e.g.,][]{Calvet+93} that they are launched from the disk surfaces similarly to the cTT and HAe winds considered in our model. For the high rates of these outflows ($\dot M_\mathrm{g,out}$\,$\sim$\,$10^{-6}$--$10^{-5}$\,$M_\sun\,\mathrm{yr}^{-1}$), Equation~\eqref{eq:amax} implies that any grains in the observed size range ($\sim$\,0.1--1\,$\micron$) that reach the base of the wind are carried away by the outflow and do not  reenter the disk. However, given that the outburst region evidently does not extend beyond $\sim$1\,au \citep[e.g.,][]{Zhu+07}, it can be expected that any such grains in fact sublimate in the intense FUOR radiation field,\footnote{For example, in the case of the archetype FU Orionis ($L$\,$\approx$\,500\,$\L_\sun$), $r_\mathrm{sub}(a=1\,\micron)$\,$\approx$\,7.6\,au (see Equation~\eqref{eq:rsub} and Table~\ref{tab:table2}).} so that the outflow remains dust free. The observed amorphous silicates are perhaps  associated with a cTT-type disk outflow that continues to operate in the colder, outer regions of the disk.

Another prediction of our model that is consistent with observational findings is an inverse correlation between the mass-averaged size of grains at the base of the wind---which  is roughly $a_\mathrm{max}(r)$ in this model---and the disk radius $r$ (see Equation~\eqref{eq:amax} and Figure~\ref{fig:fig3}). Such a trend was deduced from an analysis of the mean grain size in two large samples of protoplanetary disks \citep{Oliveira+11} and from a study of the behavior of crystalline grains in cTT and HAe systems \citep{Bouwman+08,Juhasz+10}. In the latter case, this result follows from the inferences that the mass ratio of enstatite (MgSiO$_3$) to forsterite (Mg$_2$SiO$_4$) crystals decreases with disk radius and that the typical size of enstatite grains is larger than that of the forsterite ones. The indicated radial gradient in the composition of crystalline grains also has a plausible explanation in this model. It is generally understood that forsterite forms first during the thermal annealing process, and that enstatite forms from forsterite through secondary reactions that occur faster the greater the compactness of the parent grains \citep[e.g.,][]{Bouwman+08}. In the wind transport picture, the local population of reentering grains becomes dominated, as one moves in along the disk surface, by grains that arrive from progressively smaller disk radii. In the context of this scenario, one could attribute the predominance of enstatite grains at smaller radii to the more favorable conditions for their formation in the innermost disk region where the density and kinetic temperature peak.

The inferred early establishment of the surface crystallinity level and its persistence at later times were previously interpreted in the context of turbulent disk models \citep[e.g.,][]{Ciesla09,Ciesla11}, but such explanations may not be viable if extended regions of protoplanetary disks indeed have nonturbulent surface layers. In our scenario, grains could be annealed close to the protostar and transported to the outer disk regions as soon as the disk-accretion/disk-wind flow pattern is established and the intense central radiation field is turned on, which are expected to occur early in the protostellar evolution. Furthermore, in this picture the observed crystallinity fraction would not change strongly as the mass outflow rate diminishes with increasing age because it depends primarily on the radiative properties of the central star ($L_*$, and to a lesser extent $T_*$), which evolve much more slowly. Thus, while the characteristic size of reentering grains would decrease with time (as discussed above), the fraction of annealed grains (which is determined by the dust temperature distribution along the disk surface; see Figures~\ref{fig:fig5}--\ref{fig:fig7}) would remain roughly constant since $T_\mathrm{d}$ depends only weakly on the grain size $a$ (see Equation~\eqref{eq:rsub}).

For mass outflow rates of the order of our fiducial value, the column density of the dust uplifted from the innermost disk region could be large enough to shield the disk surface farther out from the stellar UV radiation, which may preclude the launching of a vigorous disk outflow beyond $\sim$\,2\,$r_\mathrm{sub}$ (\citealt{BansKonigl12}; see also \citealt{Panoglou+12}). The existence of a disk outflow of this type has been indicated by observations of the HAe star HD~163296, in which, on the one hand, the detection of correlated optical and infrared variability points to the presence of a dusty outflow in the inner disk \citep{Ellerbroek+14}, and, on the other hand, the measurement of low turbulent velocities in the outer disk can be attributed to the shielding of that region from stellar UV photons \citep{Flaherty+15,Flaherty+17,Simon+18}. Although we do not investigate such an outflow configuration explicitly, we can conclude from our existing model results that it should have similar dust transport properties if the mass outflow rate is comparable. This follows from the observations that the characteristic size of the reentering grains is not sensitive to the value of the launch radius (see Equation~\eqref{eq:amax}) and that, for a self-similar wind, each decade in radius in the region from which grains arrive at a given location along the disk surface contributes roughly equally to the mass of the reentering dust (see Figure~\ref{fig:fig4}). A detailed calculation is, however, required to determine the extent to which the predicted crystallinity fractions change: in and of itself, the truncation of the outflow region would lead to a higher proportion of annealed grains, but this effect would be mitigated by the self-shielding of the wind.

Although we have concentrated on the application to annealed grains, our model has other noteworthy implications to the study of dust in protoplanetary systems. As summarized in Section~\ref{sec:introduction}, there is growing evidence for the presence of dust in the powerful disk outflows associated with these sources, and our calculations make it possible to estimate the mass discharge and size distribution of this dust. A rough estimate of the rate of mass outflow for the dust that remains embedded in the wind can be obtained by taking the difference between the mass discharge of uplifted dust at the base of the wind and the mass inflow rate of reentering grains: $\dot M_\mathrm{d,wind} \approx \dot M_\mathrm{d,out} - (\dot M_\mathrm{r,ann}+\dot M_\mathrm{r,amo})$. The results shown in Figure~\ref{fig:fig5} indicate that, for a disk wind with $\dot M_\mathrm{g,out}$\,=\,3.5$\times$$10^{-8}\,M_{\sun}\,\mathrm{yr}^{-1}$, $\dot M_\mathrm{d,wind}$\,$\sim$\,1.3--2.0$\times 10^{-10}\,M_{\sun}\,\mathrm{yr}^{-1}$ for cTT disks and~$\sim$\,0.8--1.3$\times 10^{-10}$\break$M_{\sun}\,\mathrm{yr}^{-1}$ for HAe systems. The size spectrum of the wind-transported grains can be estimated from the following general result of our model: for a disk extending out to $r_{+}$, all grains of size $\ge$\;$a_\mathrm{max}(r_{+})$ will either not be uplifted from, or else reenter, the disk. Thus, to a good approximation, the spectrum of grains that are carried away by the disk outflow will be truncated at that size. Using our fiducial parameter values, $a_\mathrm{max}(r_{+})$\,$\sim$\,0.1\,$(\dot M_\mathrm{g,out}/10^{-8}\,M_\sun\,\mathrm{yr}^{-1})(T_1/200\,\mathrm{K})^{1/2}$ $\micron$, implying that only sub-micron--size grains are likely to remain embedded in winds from cTT and HAe disks. 

A longstanding issue in the study of dust in protoplanetary disks has been the so-called small-grains problem: the observationally inferred persistence of small ($\la$\;1\,$\micron$) grains in the surface layers of the inner disk, which is inconsistent with theoretical expectations from simple models of grain growth and vertical settling \citep[e.g.,][]{DullemondDominik05}. One proposed resolution of this problem involves the incorporation of particle fragmentation and turbulent mixing effects, which act to replenish the population of small grains at high disk altitudes. However, unless additional conditions are realized, the required level of turbulence is rather high \citep{Zsom+11,KrijtCiesla16}---corresponding to an $\alpha$ parameter of 0.01 (see Section~\ref{subsec:diffusion})---which, as was noted in Section~\ref{sec:introduction}, is inconsistent with recent theoretical work (as well as observational findings on larger scales). A disk wind model of the type that we have considered provides a straightforward alternative explanation. In this picture, small grains remain well coupled to the gas, so they are naturally present at the surface in any dusty disk region that drives an MHD wind (see Figure~\ref{fig:fig3}). In fact, these grains might even serve as an observational signature of such a wind. Note in this connection that the nested morphology of the flowlines that comprise the modeled disk--wind system (Section~\ref{subsec:disk-wind}) implies that the gas that feeds the wind stays away from the dense midplane region, so that the small grains it carries from the ambient molecular cloud core are less susceptible to coagulation and likely maintain their original distribution.

Real protoplanetary disk outflows can be expected to be significantly more complex and varied than the idealized and highly simplified model employed in this paper. Among other factors, they are affected by the ionization structure (which determines the conductivity regime of the gas---ambipolar, Hall, or Ohm), the polarity and degree of symmetry of the magnetic field, the magnitude of the plasma $\beta$ parameter, and the contribution of external heating sources. Recent studies have begun to incorporate some of these effects and have already made significant progress \citep[e.g.,][]{Turner+14,Gressel+15,Bethune+17,Bai17}, but the models remain incomplete. Insofar as our main results depend only on a few basic parameters ($M_*$, $L_*$, $\dot M_\mathrm{g,out}$) and on the assumption that the wind structure is self-similar, they should have broad qualitative applicability; however, quantitative predictions may well require using a more elaborate scheme. Improved accuracy would also necessitate a more detailed treatment of the thermal structure of the wind.

\section{Conclusion}
\label{sec:conclusion}

Observations of protoplanetary disks have provided evidence for the presence of $\sim$\,0.1--1\,$\micron$ crystalline silicate grains in their surface layers. The mean crystalline mass fractions are estimated to be $\sim$10-20\%, and it is inferred that the degree of crystallinity is established early in the disk evolution and changes little with age. The crystalline grains could have formed by thermal annealing of amorphous dust, but they are detected on spatial scales where the dust temperature  $T_\mathrm{d}$ is much lower than the threshold value ($\sim$$10^3$\,K) for this process. Similar inferences have also been made for the protosolar disk based on data from short-period comets. One possibility---supported by indications in some sources of a decrease in the crystallinity fraction on going from the inner to the outer disk regions---is that the grains are annealed at small disk radii (where $T_\mathrm{d}$ is high) and are subsequently transported to the cold outer regions of the disk. In this paper we examine a scenario of this type, in which an MHD disk wind uplifts dust from the disk and transports it outward, with intermediate-size grains eventually succumbing to the tidal gravitational force and reentering the disk. In this picture, the uplifted grains are thermally processed by the stellar radiation, to which they become exposed after leaving the disk. In contradistinction with the previously proposed X-wind model \citep[e.g.,][]{Shu+01}, the magnetic field that drives the wind is not confined to the edge of the stellar magnetosphere, so this model is not subject to the caveat that $T_\mathrm{d}$ at the launch radius may exceed the dust sublimation temperature. We use the semianalytic, radially self-similar disk--wind solution of \citet{Teitler11}, which incorporates a nonideal-MHD accretion flow and an ideal-MHD magnetocentrifugal outflow, to describe the underlying gas dynamics. We demonstrate that collisional effects are by and large unimportant in the wind region, so that the equations of motion for the uplifted dust remain linear. We solve these equations by two independent methods: a Monte Carlo scheme and a grid-based matrix decomposition algorithm. 

We find that our model naturally accounts for many of the observational findings. A key quantity for a qualitative interpretation of the results is $a_\mathrm{max}(r)$, the maximum grain size that can be uplifted from the base of the wind at radius~$r$, which scales as $\dot M_\mathrm{g,out}T_\mathrm{g}^{1/2}(r)/M_*$ (see Equation~\eqref{eq:amax}). For characteristic parameter values of cTT and HAe systems, $a_\mathrm{max}$---which is expected to vary only weakly with $r$---lies in the observed size range of annealed grains. To a good approximation, the uplifted grains that reenter the disk are those whose size~$a$ corresponds to $a_\mathrm{max}(r)$ for some radius within the disk, and a significant fraction of them converge to a region that lies just beyond this radius (which we label $r_\mathrm{max}(a)$---it is the maximum radius from which grains of size~$a$ can be uplifted). According to this picture, the characteristic sizes of grains that are uplifted from---and then reenter---the disk are determined by the competition between the drag force that pushes the dust up (represented by the $\dot M_\mathrm{g,out}$ term in Equation~\eqref{eq:amax}) and the tidal gravitational force that pulls the grains back down (the $M_*$ term), as well as by the strength of the coupling between the gas and the dust (the $T_\mathrm{g}^{1/2}$ term): grains with $a$\,$\ll$\,$a_\mathrm{max}(r)$ are carried away by the outflow, whereas larger grains remain in the disk. In view of the fact that $\dot M_\mathrm{g,out}/M_*$ decreases with time, this picture predicts that larger grains can be uplifted and transported outward when the system is younger: this could be relevant to the origin of the 20\,$\micron$ CAI-like particle that was identified in one of the samples returned by the Stardust mission from Comet 81P/Wild~2. Furthermore, this model implies an inverse correlation between the mass-averaged size of grains at the base of the wind and the disk radius (reflecting the expectd dependence of $T_\mathrm{g}$ on $r$): this prediction, too, is consistent with observational inferences. Another implication of this scenario is to the origin of the $\la$\;1\,$\micron$ grains in the surface layers of protoplanetary disks (the so-called small-grains problem): their inferred persistence at these locations can be regarded, in the context of this model, as an observational signature of the disk wind in which they are embedded. There is observational evidence also for the dust within the outflows that are driven from the surfaces of these disks, and our model provides an estimate for the  largest size of these grains: it is roughly the maximum size of grains that can be uplifted from the disk's outer edge, which is $<$\,1\,$\micron$ for typical parameters. The mass outflow rate of wind-borne dust as a fraction of $\dot M_\mathrm{g,out}$ is found to be $\sim$0.4--0.6\% for cTT systems and~$\sim$0.2--0.4\% for HAe disks (where the listed ranges reflect different choices of the gas temperature profile; see Equation~\eqref{eq:Tg} and Table~\ref{tab:table1}). These considerations should also be relevant to other astrophysical sources where dusty disks drive outflows, notably active galactic nuclei.

The Monte Carlo scheme used for deriving the dynamical properties of the uplifted dust also enables us to determine the distribution of the maximum temperature to which reentering grains are heated after becoming exposed to the stellar radiation. By approximating all grains with $T_\mathrm{d,max}$\,$\ge$\,$T_\mathrm{ann}$ as being annealed and all those with lower maximum temperatures as remaining amorphous, we calculate the mass fraction of crystals among the reentering silicate grains at any location along the disk as well as over the entire extent of the disk, and we estimate the predicted local (resp., global) crystallinity fraction by evaluating the contribution of reentering crystalline grains to the total density (resp., total mass) of dust at the base of the wind. For disks with $r_{+}$\,=\,100\,au, the global fractions determined in this way using a conservative estimate of $T_\mathrm{ann}$ in the context of the \citet{Fabian+00} annealing law are $\sim$3--8\% for cTT disks and $\sim$7--17\% for HAe systems. These values would be somewhat higher if the $T_\mathrm{ann}$ estimates were lower, but in any case they are only lower limits on the observed crystallinity fractions because of the tendency of processed grains to accumulate near their respective convergence radii. This tendency is a consequence of the fact that the flow stagnation point for reentering grains of size $a$ is located just below the base of the wind very close to $r_\mathrm{max}(a)$, and its potential importance is indicated by the pronounced density spikes that appear at these locations in the equilibrium solution for the dust distribution in the full disk--wind model (which we derive using the grid-based algorithm; see Appendix~\ref{sec:AppB}). The main factor that determines the crystallinity fraction in this picture is the stellar luminosity, which evolves slowly with protostellar age: this could account for the inferred near-constancy of this quantity after it is established early in the disk evolution. We find that, at any given radius, the distribution of the mass of reentering grains as a function of $T_\mathrm{d,max}$ has a flattish appearance, corresponding to the flat dependence of this mass on the logarithm of the grain launch radius (which, in turn, is a reflection of the self-similarity of the wind model). As the disk radius increases, a larger fraction of the reentering grains have $T_\mathrm{d,max}$\,$<$\,$T_\mathrm{ann}$, so that, beyond a certain radius ($\sim$\,0.5\,au and $\sim$\,5\,au for the model cTT and HAe disks), the bulk of the reentering dust mass is amorphous. This behavior could account for the observationally inferred radial crystallinity gradients (which are predicted by this model to have different spatial scales in cTT and HAe systems). The wind transport scenario also appears to be consistent with the reported  radial gradient in the composition of crystalline grains (transitioning from enstatite-dominated to forsterite-dominated as $r$ increases).

The main results of our model are insensitive to the detailed structure of the disk--wind system and should be robust in that they only depend on a few basic parameters ($M_*$, $L_*$, $\dot M_\mathrm{g,out}$) and on the assumption that the flow geometry can be approximated as being self-similar. However, these expectations need to be confirmed by explicit calculations. It would also be useful to investigate the possibility that dust uplifted close to the central star shields the outer disk and wind regions from the stellar UV photons. Such shielding could affect the degree of ionization and the temperature of the upper disk layers farther out---and therefore the strength of MHD outflows from the outer portions of the disk, as well as the gas and dust temperatures in the outer wind zone---and therefore the dynamics and thermal structure of uplifted grains in that region.

\acknowledgements

We thank the referee for helpful comments. This work was supported in part by NASA ATP grant NNX13AH56G. SK acknowledges support through Hubble Fellowship Program HST-HF2-51394.002-A, provided by NASA through a grant from the Space Telescope Science Institute, which is operated by the Association of Universities for Research in Astronomy, Inc., under NASA contract NAS 5-26555.
 
\appendix
\section{Validity of Assumptions about the Dust Dynamics}
\label{sec:AppA}

Two key assumptions that underlie the dust transport equations presented in Section~\ref{subsec:dynamics} are that the dust velocity is well approximated by its time-asymptotic form and that the effects of grain--grain collisions (including, in particular, grain coagulation and fragmentation) can be neglected. In this Appendix we assess the validity of these assumptions.

\subsection{Time-Asymptotic Form of $\mathbf{v}_\mathrm{d}$}
\label{subsec:asymptotic}

\begin{figure*}[htb!]
\centering
\includegraphics[width=\textwidth]{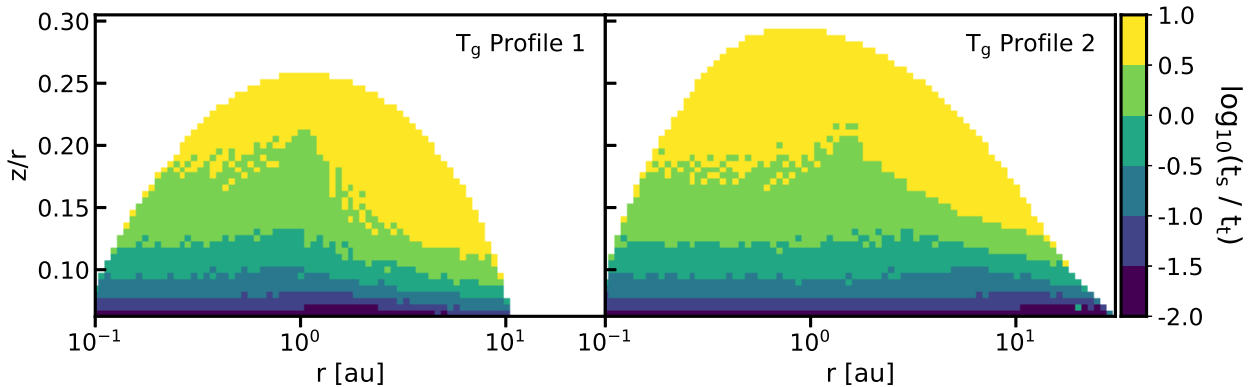}
\caption{Color-coded maps showing the spatial distribution of the ratio of the grain stopping time $t_\mathrm{s}$ within a given grid cell to its transit time $t_\mathrm{t}$ across the cell for wind-uplifted 1\,$\micron$ grains. The left and right panels correspond to $T_\mathrm{g}$ profiles~1 and~2, respectively. The terminal-velocity approximation for the grain motion is well justified in regions where $t_\mathrm{s}/t_\mathrm{t}$\,$<$\,1.}
\label{fig:fig8}
\end{figure*}

The assumption of a time-asymptotic form enabled us to formulate a steady-state problem by setting $d\mathbf{v}_\mathrm{d}/dt$\,=\,0. The e-folding time  over which an accelerated grain of size $a$ attains its terminal velocity at any given location in the wind is given by the local value of the stopping time $t_\mathrm{s}(a)$, so this approximation would be valid for any grid cell in the computational domain where $t_\mathrm{t}(a)$, the particle transit time through the cell, satisfies $t_\mathrm{s}(a)/t_\mathrm{t}(a)$\,$<$\,1. We evaluate this ratio by using the prescriptions in Sections~\ref{subsec:dynamics} and~\ref{subsec:MC}, respectively, to determine $t_\mathrm{s}(a)$ and $t_\mathrm{t}(a)$. The results for reentering grains of size $a$\,=\,1\,$\micron$ and two $T_\mathrm{g}$ profiles are presented in Figure~\ref{fig:fig8}. We find that this ratio is $<$\,1 when the values of both $z/r$ and $r$ are low ($\la$\,0.12 and $\la$\,1\,au, respectively, in the case of the fiducial model), but that $t_\mathrm{s}$ exceeds $t_\mathrm{t}$ further up and out, in regions where the high grain velocities and low gas densities have the effect of (respectively) decreasing the particle transit times and increasing their stopping times. It can, however, be seen by referring to Figure~\ref{fig:fig1} that, under the assumption that the launch locations are distributed uniformly in $\log{(r_\mathrm{l})}$ (see Equation~\eqref{eq:xi}), most of the trajectories lie entirely within the $t_\mathrm{s}/t_\mathrm{t}$\,$<$\,1 zones in Figure~\ref{fig:fig8}. An equivalent statement (see Figure~\ref{fig:fig4}) is that most of the dust mass that reenters the disk is brought in along trajectories that pass entirely within these zones. Based on these results, we consider the terminal-velocity approximation to be adequate.

\subsection{Neglect of Grain--Grain Collisions}
\label{subsec:collisions}

If collisions between grains were important in the disk outflows that we model, the problem of dust transport in the wind would be significantly more difficult. For one thing, the equations would no longer be linear, as the collision terms are proportional to the product of the number densities of the colliding particles. Another complication would arise from the need to keep track of possible transitions between particle size bins along the grains' trajectories, which themselves are size dependent. 

We assess the importance of grain--grain collisions in the following manner. Consider a grain of size $a_i$ that moves through a given grid cell in the computational domain. Its collision rate with a grain in the particle size bin $k$ is given by
\begin{equation}
C_k = n_k \, v_{i-k} \, \sigma_\mathrm{col}\,,
\label{eq:Ck}
\end{equation}
where $n_k$ is the number density of grains in the size bin $k$, $v_{i-k}$ is the relative speed between the particle of size $a_i$ and the particle of size $a_k$, and $\sigma_\mathrm{col} = \pi (a_i + a_k)^2$ is the geometrical collision cross section. Since, in our model, two grains with identical sizes move with the same velocity within a given cell, we substitute $a_k$\,=\,0.99\,$a_i$ when calculating $C_k$ for $k$\,=\,$i$. Furthermore, to take account of the fact that collisions would have a significant effect on a grain's mass (through either coagulation or fragmentation) only if they involved an interaction with a comparable (or larger) mass, we modify the form of the collision rate to
\begin{equation}
\Tilde{C}_k = \left\{ \begin{array}{ll}
C_k \quad &\text{if} \ (a_k/a_i)^3 > 1\;,\\
(a_k/a_i)^3 \, C_k\quad  &\text{if} \; (a_k/a_i)^3 \leq 1\;.
\end{array} \right.
\label{eq:TCk}
\end{equation}
In this formulation, the collisions are weighted according to their effect on the grain's mass. Neglecting the possibility of multiple major collisions, the probability for the given grain to undergo significant mass alteration during its transit time $t_\mathrm{t}$ through the cell is given by
\begin{equation}
\Tilde{P}_\mathrm{col} = 1 - e^{-\Tilde{C}_\mathrm{tot} t_\mathrm{t}}\, ,
\label{eq:Pcol}
\end{equation}
where
\begin{equation}
\Tilde{C}_{tot} = \sum_k\Tilde{C}_k
\label{eq:Ctot}
\end{equation}
\citep[cf.][]{KrijtCiesla16}.

\begin{figure*}[htb!]
\centering
\includegraphics[width=\textwidth]{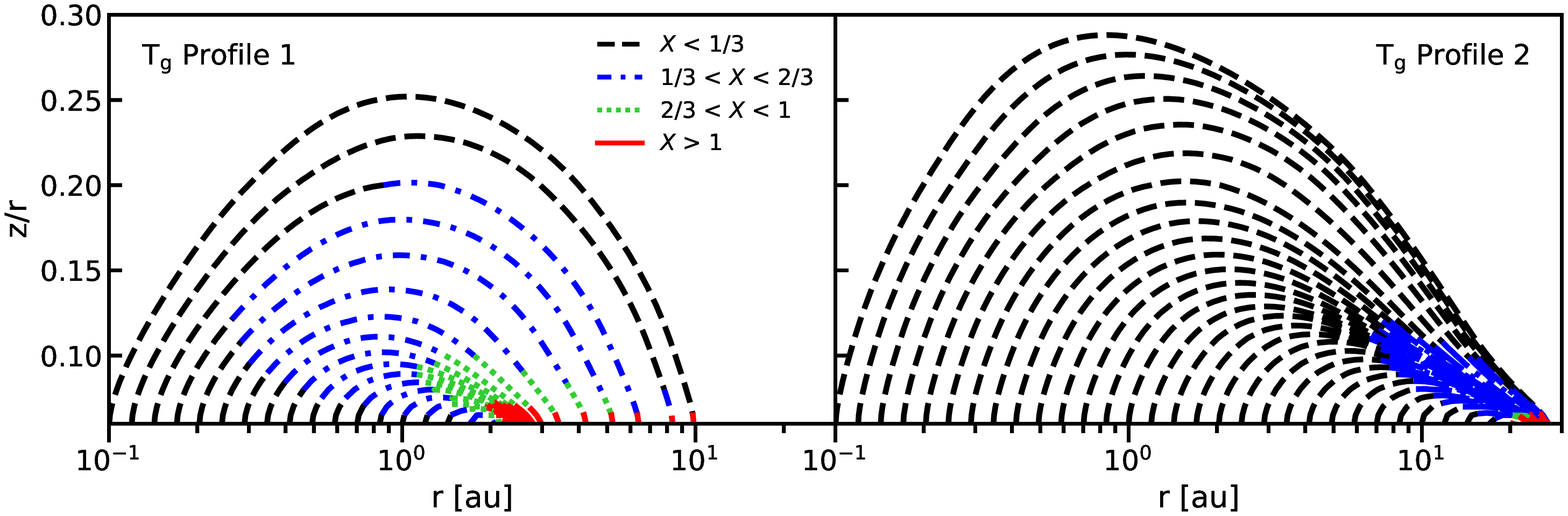}
\caption{Values of the cumulative collision-probability parameter $X$ (defined in Section~\ref{subsec:collisions}) along the 1\,$\micron$ grain trajectories for gas temperature profiles~1 and~2. Mass-altering grain--grain collisions are unimportant in regions where $X$\,$<$\,1.}
\label{fig:fig9}
\end{figure*}

Defining $dX$\,$\equiv$\,$\Tilde{C}_\mathrm{tot}\,t_\mathrm{t}$, we can evaluate the cumulative contribution to $P_\mathrm{col}$ from the grid cells that the given grain traverses by integrating along the particle's trajectory, starting at the grain's launch point. The results for $a_i$\,=\,1\,$\micron$ grains are shown in Figure~\ref{fig:fig9}, where the two panels illustrate the dependence on the magnitude of the parameter $T_1$ in the expression (Equation~\eqref{eq:Tg}) for the gas temperature profile. Collisional effects can be regarded as being unimportant everywhere along the grain's path where $X$ remains $<$\,1, and the figure demonstrates that this is the case everywhere except in a narrow region at the bottom of the grid where the trajectories of the reentering grains approach the disk surface. It is also seen that the higher value of $T_1$ that characterizes Profile~2 has the effect of reducing the values of $X$, which can be attributed to the fact that the 1\,$\micron$ grains are better coupled to the gas in this case (given that $t_\mathrm{s}$\,$\propto$\,$T_1^{-1/2}$; see Section~\ref{subsec:dynamics}) and therefore have a lower collision rate with the well-coupled small particles that dominate the grain population.

Our main conclusion from these results is that, though collisions may start to affect the distribution of reentering grains when they approach the disk surface, they should have a negligible effect on the dynamics of uplifted grains while they travel in the wind.

\section{Grid-Based Solutions}
\label{sec:AppB}
\setcounter{equation}{0}

In Section~\ref{sec:results} we employ a Monte Carlo scheme to determine the grain distribution in the wind zone since this method also makes it possible to derive the temperature distribution of the reentering grains. However, the steady-state grain distribution can alternatively be obtained through a numerical, grid-based solution of the transport equations (Section~\ref{subsec:formulation}). The grid-based approach is useful in that it makes it possible to check the Monte Carlo results (Section~\ref{subsec:wind}) and, when extended to also cover the disk (Section~\ref{subsec:full}), to gain an insight into the long-term evolution of the reentering grains (see Section~\ref{sec:discussion}). The equilibrium solution for the dust distribution in the full disk--wind model is, however, incomplete in that it does not incorporate grain--grain collisions and particle diffusion, which can be important in regions of high particle density. It is nonetheless possible to assess the effect of the latter process by incorporating phenomenological diffusivity terms into the dust transport equations (Section~\ref{subsec:diffusion}).

\subsection{Formulation}
\label{subsec:formulation}

The transport equations are solved on a spatial grid that covers the modeled region. The grid is divided by vertical boundaries separating columns of cells and by boundaries that correspond to the self-similarity surfaces (i.e., surfaces of constant polar angle). Thus, each spatial cell is a trapezoid with vertical sides and with tilted top and bottom surfaces that are not quite parallel.  We assume that dust grains can move from a given spatial cell to at most four other cells: the cells immediately above and below it, and the cells that lie just inward and outward of it in the (spherical) radial direction. We further assume that the grains move at the terminal velocity calculated at the center of the cell: based on the results of Section~\ref{subsec:asymptotic}, this assumption should be valid inside the disk and at sufficiently low altitudes within the wind.

The steady-state evolution of the dust number density is given by Equation~\eqref{eq:nd}. In its grid-based form, it is transformed into the following set of equations for the number densities $n(i,j,k)$ of grains of size $k$ in cells identified by the labels of their column ($i$) and row ($j$):
\begin{equation}
n(i,j,k)|v_{\mathrm{d},r}(i,j,k)|/\Delta r(i,j) + n(i,j,k)|v_{\mathrm{d},z}(i,j,k)|/\Delta z(i,j) = I(i,j,k)\,,
\label{eq:translineq}
\end{equation}
where $v_{\mathrm{d},r}$ and $v_{\mathrm{d},z}$ are given, respectively, by Equations~\eqref{eq:vr} and~\eqref{eq:vz},\footnote{Note that Equation~\eqref{eq:translineq} is constrained not to admit a downward vertical loss term ($v_{\mathrm{d},z}$\,$<$\,0) for $j$\,=\,1 (the midplane).} $\Delta r$ (resp., $\Delta z$) is the (cylindrical) radial (resp., vertical) thickness of the cell, and $I$ is the influx of particles from neighboring cells. We calculate the dust velocity using the dynamical model described in Section~\ref{sec:model}. However, in evaluating the stopping time $t_\mathrm{s}$ in Equations~\eqref{eq:vr} and~\eqref{eq:vz}, we assume that $T_\mathrm{g}\propto$\,$r^{-1}$ (the self-similarity scaling) along the midplane and that the gas remains vertically isothermal: this corresponds to a temperature profile of the form given by Equation~\eqref{eq:Tg}, with $T_1$\,=\,250\,K, $\gamma$\,=\,0, and $\delta$\,=\,1.

Except for cells along the boundaries of the computational domain, there are four neighboring cells that can contribute influx terms:
\begin{eqnarray}
    I_{\mathrm{above}}(i,j,k) & =& -\left[n(i,j+1,k)v_{\mathrm{d},z}(i,j+1,k)/\Delta z(i,j+1)\right]\left [V(i,j+1)/V(i,j) \right]\nonumber\\
    I_{\mathrm{below}}(i,j,k) & = &\left [(i,j-1,k)v_{\mathrm{d},z}(i,j-1,k)/\Delta z(i,j-1)\right]\left[ V(i,j-1)/V(i,j)\right ] \nonumber\\
    I_{\mathrm{left}}(i,j,k) & = &\left [(n(i-1,j,k)v_{\mathrm{d},r}(i-1,j,k)/\Delta r(i-1,j)\right]\left [V(i-1,j)/V(i,j)\right ] \nonumber \\
    I_{\mathrm{right}}(i,j,k) & = &-\left[n(i+1,j,k)v_{\mathrm{d},r}(i+1,j,k)/\Delta r(i+1,j)\right]\left [V(i+1,j)/V(i,j)\right ] \,,
\label{eq:I}
\end{eqnarray}
where $V$ is the cell volume and the labels ``left'' and ``right'' are shorthand, respectively, for ``radially inward'' and ``radially outward.'' In all cases, if the expression for an influx term is negative, it is set equal to zero. The minus signs in the definitions of $I_{\mathrm{above}}$ and $I_{\mathrm{right}}$ ensure positive influxes for dust moving down from above (negative $v_{\mathrm{d},z}(i,j+1,k)$) and inward from the right (negative $v_{\mathrm{d},r}(i+1,j,k)$).

\subsection{Wind-Zone Solution}
\label{subsec:wind}

Here we apply a grid-based approach to the same region considered in the Monte Carlo solutions presented in Section~\ref{sec:results}. Throughout this region, $v_{\mathrm{d},r} > 0$ for all grains, so $I_{\mathrm{right}}(i,j,k) = 0$ in all cases. Our boundary conditions are that dust with the size distribution and density normalization adopted in Section~\ref{subsec:dynamics} can enter along the bottom edge of the model region, but that no entry is permitted along the inner and outer radial boundaries or along the top edge of the computational domain. For any given cell at the base of the wind, a grain in a size bin $k$ can enter the wind only if it has $v_{\mathrm{d},z}$\,$>$\,0 at the bottom of that cell.

We use a grid of 100$\times$19 (radial$\times$vertical) cells extending radially between 0.1 and 100\,au and vertically between the base of the wind and
$z/r$\,=\,0.19 (corresponding to an altitude of 6 scale heights), so that for each grain size we have a system of 1900 linear equations. This is, however, an extremely sparse system, with each equation having at most four terms. (Since $I_{\mathrm{right}}(i,j,k) = 0$, the equation for $n(i,j,k)$ includes an outflow term that depends on $n(i,j,k)$ and at most 3 influx terms, which depend on $n(i,j+1,k)$, $n(i,j-1,k)$, and $n(i-1,j,k)$.) In fact, most equations have only three terms, since (for a grain in bin size $k$) at most one cell in any given column has $v_{\mathrm{d},z}(i,j+1,k)$\,$<$\,0 and $v_{\mathrm{d},z}(i,j-1,k)$\,$>$\,0. We solve this system using a band diagonal matrix LU decomposition algorithm from \citet{Press+96}.

\begin{figure}[htb!]
\includegraphics[width=\textwidth]{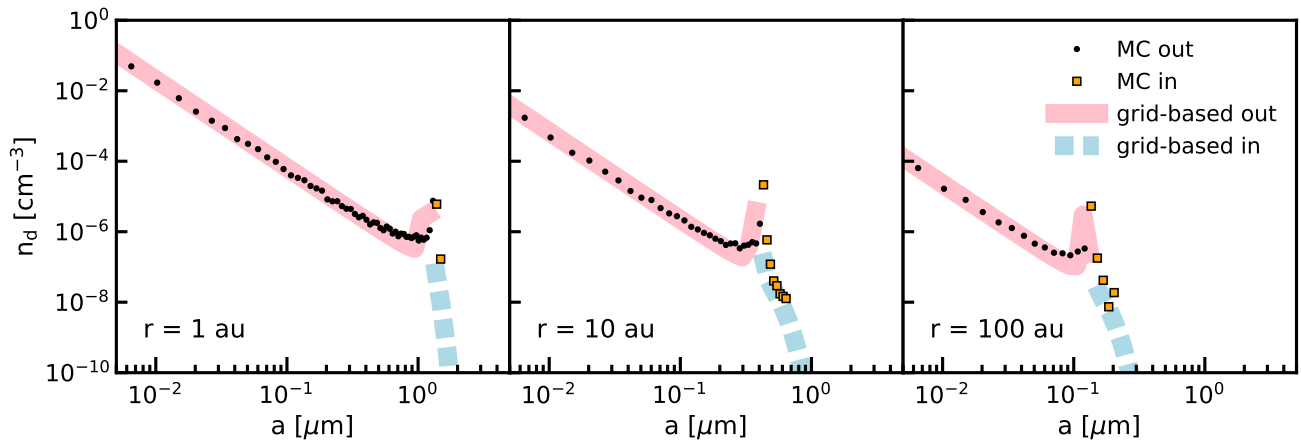}
\caption{Comparison between the grid-based solution for the wind zone (Section~\ref{subsec:wind}) and the corresponding solution obtained using the Monte Carlo method described in Section~\ref{subsec:MC}. The dust number density distribution at the base of the wind is plotted as a function of the grain size and shown for three different radii. The densities were calculated using the gas temperature profile specified in Section~\ref{subsec:formulation}. The black circles and orange squares represent outgoing and reentering grains, respectively, in the Monte Carlo solution, whereas the corresponding results for the grid-based solution are shown by solid pink and dashed light-blue lines.}
\label{fig:fig10}
\end{figure}

The results obtained from this solution are compared with those derived using the Monte Carlo method in Figure~\ref{fig:fig10}, which presents the grain number density distribution at the base of the wind for the same three radii selected in previous figures: it is seen that there is an excellent agreement between the two solutions. 

\subsection{Solution over Entire Disk--Wind Domain}
\label{subsec:full}

Our underlying MHD solution \citep{Teitler11} describes a continuous flow configuration that includes both the disk and the wind. It is therefore possible to solve the dust transport equations over the entire accretion--outflow region rather than just over the wind zone as is done in Section~\ref{subsec:wind}. To that end, we enlarge the computational domain and change the boundary conditions. We consider the same radial interval as in the wind-only solution (0.1$-$100\,au) but extend the vertical interval from the base of the wind to the midplane (so that the entire $z/r$ range from 0 to 0.19 is included). Our grid now comprises 100$\times$48 cells, resulting in a (still extremely sparse) system of 4800 linear equations for each particle size. The modified boundary conditions reflect the different setup: particles are only allowed to enter through the outer radial boundary of the computational domain, and, for any given cell along that boundary, a grain in a size bin $k$ can enter the disk only if it has $v_{\mathrm{d},r}$\,$<$\,0 at the cell's outer edge. The density distribution of the incoming dust is the same as in the wind-only implementation: the size distribution is given by Equation~\eqref{eq:dnda} and satisfies $\rho_\mathrm{d}$\,=\,0.01\,$\rho_\mathrm{g}$.

\subsection{Effect of Particle Diffusion}
\label{subsec:diffusion}

The dust distribution for the full disk--wind configuration exhibits unrealistically large density spikes. As pointed out in Section~\ref{sec:discussion}, the incorporation of grain--gain collisions and particle diffusion effects could be expected to mitigate the density growth. If one were to incorporate collisional effects into our model, the equations would become quadratic. Using the same setup, we would have $4800K$ equations to solve (with $K$ being the number of grain size bins), each depending on at most $K+4$ density values (i.e., the system would again be very sparse). There is, however, no general method for obtaining a solution of such a large system of nonlinear equations. In contrast, the inclusion of particle diffusion terms keeps the transport equations linear, so the model can be practicably extended to examine the possible effect of this process.

To incorporate particle diffusion, we follow \citeauthor{Miyake+16} (\citeyear{Miyake+16}; see also \citealt{TakeuchiLin02}) and generalize the particle flux in Equation~\eqref{eq:nd}  by adding to the advective term ($n_\mathrm{d}\mathbf{v}_\mathrm{d}$) a diffusive term of the form
\begin{equation}
\mathbf{J}_\mathrm{d} = - \frac{n_\mathrm{g} \nu_\mathrm{g}}{\mathrm{Sc}}\; \mathbf{\nabla} \left (\frac{n_\mathrm{d}}{n_\mathrm{g}}\right ) ,
\label{eq:Jd}
\end{equation}
where $\nu_\mathrm{g}$ is the effective (turbulent) viscosity of the gas and Sc is the Schmidt number (the gas-to-particle diffusivity ratio), which is approximated as  Sc\,$\approx$\,$1+(rt_\mathrm{s}/v_\mathrm{K})^2$ \citep{YoudinLithwick07}. The viscosity is written in the common ``$\alpha$ parameter'' representation as $\nu_\mathrm{g} = \alpha c_\mathrm{s}^2 r /v_\mathrm{K}$.

We discretize the equation for $\mathbf{J}_\mathrm{d}$ as follows. For any two cells 1 and 2 that undergo diffusive interaction, we define the fluxes
\begin{equation}
J_{1-2} = \left(n_{\mathrm{g},1}D_{\mathrm{d},1}+n_{\mathrm{g},2}D_{\mathrm{d},2}\right)\left(n_{\mathrm{d},1}/n_{\mathrm{g},1}\right)/(2d_{12})
\label{eq:diffeqdisc1}
\end{equation}
and
\begin{equation}
J_{2-1} = \left(n_{\mathrm{g},1}D_{\mathrm{d},1}+n_{\mathrm{g},2}D_{\mathrm{d},2}\right)\left(n_{\mathrm{d},2}/n_{\mathrm{g},2}\right)/(2d_{12})\,,
\label{eq:diffeqdisc2}
\end{equation}
where $d_{12}$ is the distance between the two cell centers and $D_\mathrm{d}$\,$\equiv$\,$\nu_\mathrm{g}/\mathrm{Sc}$. The diffusive flux affecting the dust density in cell \#1 is given by $(V_2/V_1)J_{2-1}$\,-\,$J_{1-2}$ (where $V_1$ and $V_2$ are the respective cell volumes), with the corresponding expression for cell \#2 being $(V_1/V_2)J_{1-2}$\,-\,$J_{2-1}$. This prescription makes it possible to evaluate the diffusive flux without having to know its direction in advance. For cells located at the lower, upper, and outer radial boundaries, we evaluate the flux using a ``mirror'' boundary condition: the flux into (resp., out of) the cell through the boundary is the same as the flux out of (resp., into) the cell through the opposite edge. However, along the inner edge of the computational domain, we require instead that there be no diffusion of dust into the cell through the boundary.


\begin{thebibliography}{84}
\expandafter\ifx\csname natexlab\endcsname\relax\def\natexlab#1{#1}\fi

\bibitem[{{Apai} {et~al.}(2005){Apai}, {Pascucci}, {Bouwman}, {Natta},
  {Henning}, \& {Dullemond}}]{Apai+05}
{Apai}, D., {Pascucci}, I., {Bouwman}, J., {et~al.} 2005, Science, 310, 834

\bibitem[{{Bai}(2017)}]{Bai17}
{Bai}, X.-N. 2017, \apj, 845, 75

\bibitem[{{Bans} \& {K{\"o}nigl}(2012)}]{BansKonigl12}
{Bans}, A., \& {K{\"o}nigl}, A. 2012, \apj, 758, 100

\bibitem[{{Banzatti} {et~al.}(2019){Banzatti}, {Pascucci}, {Edwards}, {Fang},
  {Gorti}, \& {Flock}}]{Banzatti+19}
{Banzatti}, A., {Pascucci}, I., {Edwards}, S., {et~al.} 2019, \apj, 870, 76

\bibitem[{{B{\'e}thune} {et~al.}(2017){B{\'e}thune}, {Lesur}, \&
  {Ferreira}}]{Bethune+17}
{B{\'e}thune}, W., {Lesur}, G., \& {Ferreira}, J. 2017, \aap, 600, A75

\bibitem[{{Bjerkeli} {et~al.}(2016){Bjerkeli}, {van der Wiel}, {Harsono},
  {Ramsey}, \& {J{\o}rgensen}}]{Bjerkeli+16}
{Bjerkeli}, P., {van der Wiel}, M.~H.~D., {Harsono}, D., {Ramsey}, J.~P., \&
  {J{\o}rgensen}, J.~K. 2016, \nat, 540, 406

\bibitem[{{Bjorkman} \& {Wood}(2001)}]{BjorkmanWood01}
{Bjorkman}, J.~E., \& {Wood}, K. 2001, \apj, 554, 615

\bibitem[{{Blandford} \& {Payne}(1982)}]{BlandfordPayne82}
{Blandford}, R.~D., \& {Payne}, D.~G. 1982, \mnras, 199, 883

\bibitem[{{Bouwman} {et~al.}(2008){Bouwman}, {Henning}, {Hillenbrand}, {Meyer},
  {Pascucci}, {Carpenter}, {Hines}, {Kim}, {Silverstone}, {Hollenbach}, \&
  {Wolf}}]{Bouwman+08}
{Bouwman}, J., {Henning}, T., {Hillenbrand}, L.~A., {et~al.} 2008, \apj, 683,
  479

\bibitem[{{Brownlee} {et~al.}(2006){Brownlee}, {Tsou}, {Al{\'e}on},
  {Alexander}, {Araki}, {Bajt}, {Baratta}, {Bastien}, {Bland}, {Bleuet},
  {Borg}, {Bradley}, {Brearley}, {Brenker}, {Brennan}, {Bridges}, {Browning},
  {Brucato}, {Bullock}, {Burchell}, {Busemann}, {Butterworth}, {Chaussidon},
  {Cheuvront}, {Chi}, {Cintala}, {Clark}, {Clemett}, {Cody}, {Colangeli},
  {Cooper}, {Cordier}, {Daghlian}, {Dai}, {D'Hendecourt}, {Djouadi},
  {Dominguez}, {Duxbury}, {Dworkin}, {Ebel}, {Economou}, {Fakra}, {Fairey},
  {Fallon}, {Ferrini}, {Ferroir}, {Fleckenstein}, {Floss}, {Flynn}, {Franchi},
  {Fries}, {Gainsforth}, {Gallien}, {Genge}, {Gilles}, {Gillet}, {Gilmour},
  {Glavin}, {Gounelle}, {Grady}, {Graham}, {Grant}, {Green}, {Grossemy},
  {Grossman}, {Grossman}, {Guan}, {Hagiya}, {Harvey}, {Heck}, {Herzog},
  {Hoppe}, {H{\"o}rz}, {Huth}, {Hutcheon}, {Ignatyev}, {Ishii}, {Ito}, {Jacob},
  {Jacobsen}, {Jacobsen}, {Jones}, {Joswiak}, {Jurewicz}, {Kearsley}, {Keller},
  {Khodja}, {Kilcoyne}, {Kissel}, {Krot}, {Langenhorst}, {Lanzirotti}, {Le},
  {Leshin}, {Leitner}, {Lemelle}, {Leroux}, {Liu}, {Luening}, {Lyon},
  {MacPherson}, {Marcus}, {Marhas}, {Marty}, {Matrajt}, {McKeegan}, {Meibom},
  {Mennella}, {Messenger}, {Messenger}, {Mikouchi}, {Mostefaoui}, {Nakamura},
  {Nakano}, {Newville}, {Nittler}, {Ohnishi}, {Ohsumi}, {Okudaira},
  {Papanastassiou}, {Palma}, {Palumbo}, {Pepin}, {Perkins}, {Perronnet},
  {Pianetta}, {Rao}, {Rietmeijer}, {Robert}, {Rost}, {Rotundi}, {Ryan},
  {Sandford}, {Schwandt}, {See}, {Schlutter}, {Sheffield-Parker},
  {Simionovici}, {Simon}, {Sitnitsky}, {Snead}, {Spencer}, {Stadermann},
  {Steele}, {Stephan}, {Stroud}, {Susini}, {Sutton}, {Suzuki}, {Taheri},
  {Taylor}, {Teslich}, {Tomeoka}, {Tomioka}, {Toppani},
  {Trigo-Rodr{\'{\i}}guez}, {Troadec}, {Tsuchiyama}, {Tuzzolino}, {Tyliszczak},
  {Uesugi}, {Velbel}, {Vellenga}, {Vicenzi}, {Vincze}, {Warren}, {Weber},
  {Weisberg}, {Westphal}, {Wirick}, {Wooden}, {Wopenka}, {Wozniakiewicz},
  {Wright}, {Yabuta}, {Yano}, {Young}, {Zare}, {Zega}, {Ziegler}, {Zimmerman},
  {Zinner}, \& {Zolensky}}]{Brownlee+06}
{Brownlee}, D., {Tsou}, P., {Al{\'e}on}, J., {et~al.} 2006, Science, 314, 1711

\bibitem[{{Bruderer} {et~al.}(2012){Bruderer}, {van Dishoeck}, {Doty}, \&
  {Herczeg}}]{Bruderer+12}
{Bruderer}, S., {van Dishoeck}, E.~F., {Doty}, S.~D., \& {Herczeg}, G.~J. 2012,
  \aap, 541, A91

\bibitem[{{Calvet} {et~al.}(1993){Calvet}, {Hartmann}, \& {Kenyon}}]{Calvet+93}
{Calvet}, N., {Hartmann}, L., \& {Kenyon}, S.~J. 1993, \apj, 402, 623

\bibitem[{{Caratti o Garatti} {et~al.}(2012){Caratti o Garatti}, {Garcia
  Lopez}, {Antoniucci}, {Nisini}, {Giannini}, {Eisl{\"o}ffel}, {Ray},
  {Lorenzetti}, \& {Cabrit}}]{Caratti+12}
{Caratti o Garatti}, A., {Garcia Lopez}, R., {Antoniucci}, S., {et~al.} 2012,
  \aap, 538, A64

\bibitem[{{Chiang} \& {Goldreich}(1997)}]{ChiangGoldreich97}
{Chiang}, E.~I., \& {Goldreich}, P. 1997, \apj, 490, 368

\bibitem[{{Ciesla}(2007)}]{Ciesla07}
{Ciesla}, F.~J. 2007, Science, 318, 613

\bibitem[{{Ciesla}(2009)}]{Ciesla09}
---. 2009, Meteoritics and Planetary Science, 44, 1663

\bibitem[{{Ciesla}(2010{\natexlab{a}})}]{Ciesla10a}
---. 2010{\natexlab{a}}, \apj, 723, 514

\bibitem[{{Ciesla}(2010{\natexlab{b}})}]{Ciesla10b}
---. 2010{\natexlab{b}}, \icarus, 208, 455

\bibitem[{{Ciesla}(2011)}]{Ciesla11}
---. 2011, \apj, 740, 9

\bibitem[{{Desch} {et~al.}(2010){Desch}, {Morris}, {Connolly}, \&
  {Boss}}]{Desch+10}
{Desch}, S.~J., {Morris}, M.~A., {Connolly}, Jr., H.~C., \& {Boss}, A.~P. 2010,
  \apj, 725, 692

\bibitem[{{Draine} \& {Salpeter}(1979)}]{DraineSalpeter79}
{Draine}, B.~T., \& {Salpeter}, E.~E. 1979, \apj, 231, 77

\bibitem[{{Dullemond} \& {Dominik}(2005)}]{DullemondDominik05}
{Dullemond}, C.~P., \& {Dominik}, C. 2005, \aap, 434, 971

\bibitem[{{Ellerbroek} {et~al.}(2014){Ellerbroek}, {Podio}, {Dougados},
  {Cabrit}, {Sitko}, {Sana}, {Kaper}, {de Koter}, {Klaassen}, {Mulders},
  {Mendigut{\'{\i}}a}, {Grady}, {Grankin}, {van Winckel}, {Bacciotti},
  {Russell}, {Lynch}, {Hammel}, {Beerman}, {Day}, {Huelsman}, {Werren},
  {Henden}, \& {Grindlay}}]{Ellerbroek+14}
{Ellerbroek}, L.~E., {Podio}, L., {Dougados}, C., {et~al.} 2014, \aap, 563, A87

\bibitem[{{Fabian} {et~al.}(2000){Fabian}, {J{\"a}ger}, {Henning}, {Dorschner},
  \& {Mutschke}}]{Fabian+00}
{Fabian}, D., {J{\"a}ger}, C., {Henning}, T., {Dorschner}, J., \& {Mutschke},
  H. 2000, \aap, 364, 282

\bibitem[{{Fang} {et~al.}(2018){Fang}, {Pascucci}, {Edwards}, {Gorti},
  {Banzatti}, {Flock}, {Hartigan}, {Herczeg}, \& {Dupree}}]{Fang+18}
{Fang}, M., {Pascucci}, I., {Edwards}, S., {et~al.} 2018, \apj, 868, 28

\bibitem[{{Ferreira} {et~al.}(2006){Ferreira}, {Dougados}, \&
  {Cabrit}}]{Ferreira+06}
{Ferreira}, J., {Dougados}, C., \& {Cabrit}, S. 2006, \aap, 453, 785

\bibitem[{{Flaherty} {et~al.}(2015){Flaherty}, {Hughes}, {Rosenfeld},
  {Andrews}, {Chiang}, {Simon}, {Kerzner}, \& {Wilner}}]{Flaherty+15}
{Flaherty}, K.~M., {Hughes}, A.~M., {Rosenfeld}, K.~A., {et~al.} 2015, \apj,
  813, 99

\bibitem[{{Flaherty} {et~al.}(2017){Flaherty}, {Hughes}, {Rose}, {Simon}, {Qi},
  {Andrews}, {K{\'o}sp{\'a}l}, {Wilner}, {Chiang}, {Armitage}, \&
  {Bai}}]{Flaherty+17}
{Flaherty}, K.~M., {Hughes}, A.~M., {Rose}, S.~C., {et~al.} 2017, \apj, 843,
  150

\bibitem[{{Frank} {et~al.}(2014){Frank}, {Ray}, {Cabrit}, {Hartigan}, {Arce},
  {Bacciotti}, {Bally}, {Benisty}, {Eisl{\"o}ffel}, {G{\"u}del}, {Lebedev},
  {Nisini}, \& {Raga}}]{Frank+14}
{Frank}, A., {Ray}, T.~P., {Cabrit}, S., {et~al.} 2014, in Protostars and
  Planets VI, ed. {H. Beuther et al.} (Tucson, AZ: Univ. Arizona Press), 451

\bibitem[{{Gail}(1998)}]{Gail98}
{Gail}, H.-P. 1998, \aap, 332, 1099

\bibitem[{{Gail}(2004)}]{Gail04}
---. 2004, \aap, 413, 571

\bibitem[{{Glassgold} {et~al.}(2004){Glassgold}, {Najita}, \&
  {Igea}}]{Glassgold+04}
{Glassgold}, A.~E., {Najita}, J., \& {Igea}, J. 2004, \apj, 615, 972

\bibitem[{{Green} {et~al.}(2006){Green}, {Hartmann}, {Calvet}, {Watson},
  {Ibrahimov}, {Furlan}, {Sargent}, \& {Forrest}}]{Green+06}
{Green}, J.~D., {Hartmann}, L., {Calvet}, N., {et~al.} 2006, \apj, 648, 1099

\bibitem[{{Gressel} {et~al.}(2015){Gressel}, {Turner}, {Nelson}, \&
  {McNally}}]{Gressel+15}
{Gressel}, O., {Turner}, N.~J., {Nelson}, R.~P., \& {McNally}, C.~P. 2015,
  \apj, 801, 84

\bibitem[{{Harker} \& {Desch}(2002)}]{HarkerDesch02}
{Harker}, D.~E., \& {Desch}, S.~J. 2002, \apjl, 565, L109

\bibitem[{{Harker} {et~al.}(2007){Harker}, {Woodward}, {Wooden}, {Fisher}, \&
  {Trujillo}}]{Harker+07}
{Harker}, D.~E., {Woodward}, C.~E., {Wooden}, D.~H., {Fisher}, R.~S., \&
  {Trujillo}, C.~A. 2007, \icarus, 191, 432

\bibitem[{{Hartmann} \& {Kenyon}(1996)}]{HartmannKenyon96}
{Hartmann}, L., \& {Kenyon}, S.~J. 1996, \araa, 34, 207

\bibitem[{{Hill} {et~al.}(2001){Hill}, {Grady}, {Nuth}, {Hallenbeck}, \&
  {Sitko}}]{Hill+01}
{Hill}, H.~G.~M., {Grady}, C.~A., {Nuth}, III, J.~A., {Hallenbeck}, S.~L., \&
  {Sitko}, M.~L. 2001, Proceedings of the National Academy of Science, 98, 2182

\bibitem[{{Hirota} {et~al.}(2017){Hirota}, {Machida}, {Matsushita}, {Motogi},
  {Matsumoto}, {Kim}, {Burns}, \& {Honma}}]{Hirota+17}
{Hirota}, T., {Machida}, M.~N., {Matsushita}, Y., {et~al.} 2017, Nature
  Astronomy, 1, 0146

\bibitem[{{Juh{\'a}sz} {et~al.}(2010){Juh{\'a}sz}, {Bouwman}, {Henning},
  {Acke}, {van den Ancker}, {Meeus}, {Dominik}, {Min}, {Tielens}, \&
  {Waters}}]{Juhasz+10}
{Juh{\'a}sz}, A., {Bouwman}, J., {Henning}, T., {et~al.} 2010, \apj, 721, 431

\bibitem[{{Klaassen} {et~al.}(2016){Klaassen}, {Mottram}, {Maud}, \&
  {Juhasz}}]{Klaassen+16}
{Klaassen}, P.~D., {Mottram}, J.~C., {Maud}, L.~T., \& {Juhasz}, A. 2016,
  \mnras, 460, 627

\bibitem[{{Kobayashi} {et~al.}(2011){Kobayashi}, {Kimura}, {Watanabe},
  {Yamamoto}, \& {M{\"u}ller}}]{Kobayashi+11}
{Kobayashi}, H., {Kimura}, H., {Watanabe}, S.-i., {Yamamoto}, T., \&
  {M{\"u}ller}, S. 2011, Earth, Planets, and Space, 63, 1067

\bibitem[{{K{\"o}nigl} \& {Pudritz}(2000)}]{KoniglPudritz00}
{K{\"o}nigl}, A., \& {Pudritz}, R.~E. 2000, in Protostars and Planets IV, ed.
  V.~Mannings, A.~Boss, \& S.~Russell (Tucson, AZ: Univ. Arizona Press), 759

\bibitem[{{K{\"o}nigl} \& {Salmeron}(2011)}]{KoniglSalmeron11}
{K{\"o}nigl}, A., \& {Salmeron}, R. 2011, in Physical Processes in
  Circumstellar Disks around Young Stars, ed. P.~J.~V. {Garcia} (Chicago, IL:
  Univ. Chicago Press), 283

\bibitem[{{Krijt} \& {Ciesla}(2016)}]{KrijtCiesla16}
{Krijt}, S., \& {Ciesla}, F.~J. 2016, \apj, 822, 111

\bibitem[{{Kwon} {et~al.}(2011){Kwon}, {Looney}, \& {Mundy}}]{Kwon+11}
{Kwon}, W., {Looney}, L.~W., \& {Mundy}, L.~G. 2011, \apj, 741, 3

\bibitem[{{Liffman} \& {Brown}(1995)}]{LiffmanBrown95}
{Liffman}, K., \& {Brown}, M. 1995, \icarus, 116, 275

\bibitem[{{Lucy}(1999)}]{Lucy99}
{Lucy}, L.~B. 1999, \aap, 344, 282

\bibitem[{{McKeegan} {et~al.}(2006){McKeegan}, {Al{\'e}on}, {Bradley},
  {Brownlee}, {Busemann}, {Butterworth}, {Chaussidon}, {Fallon}, {Floss},
  {Gilmour}, {Gounelle}, {Graham}, {Guan}, {Heck}, {Hoppe}, {Hutcheon}, {Huth},
  {Ishii}, {Ito}, {Jacobsen}, {Kearsley}, {Leshin}, {Liu}, {Lyon}, {Marhas},
  {Marty}, {Matrajt}, {Meibom}, {Messenger}, {Mostefaoui}, {Mukhopadhyay},
  {Nakamura-Messenger}, {Nittler}, {Palma}, {Pepin}, {Papanastassiou},
  {Robert}, {Schlutter}, {Snead}, {Stadermann}, {Stroud}, {Tsou}, {Westphal},
  {Young}, {Ziegler}, {Zimmermann}, \& {Zinner}}]{McKeegan+06}
{McKeegan}, K.~D., {Al{\'e}on}, J., {Bradley}, J., {et~al.} 2006, Science, 314,
  1724

\bibitem[{{Miyake} {et~al.}(2016){Miyake}, {Suzuki}, \& {Inutsuka}}]{Miyake+16}
{Miyake}, T., {Suzuki}, T.~K., \& {Inutsuka}, S.-i. 2016, \apj, 821, 3

\bibitem[{{Muzerolle} {et~al.}(2003){Muzerolle}, {Calvet}, {Hartmann}, \&
  {D'Alessio}}]{Muzerolle+03}
{Muzerolle}, J., {Calvet}, N., {Hartmann}, L., \& {D'Alessio}, P. 2003, \apjl,
  597, L149

\bibitem[{{Nomura} \& {Millar}(2005)}]{NomuraMillar05}
{Nomura}, H., \& {Millar}, T.~J. 2005, \aap, 438, 923

\bibitem[{{Oliveira} {et~al.}(2011){Oliveira}, {Olofsson}, {Pontoppidan}, {van
  Dishoeck}, {Augereau}, \& {Mer{\'{\i}}n}}]{Oliveira+11}
{Oliveira}, I., {Olofsson}, J., {Pontoppidan}, K.~M., {et~al.} 2011, \apj, 734,
  51

\bibitem[{{Panoglou} {et~al.}(2012){Panoglou}, {Cabrit}, {Pineau Des
  For{\^e}ts}, {Garcia}, {Ferreira}, \& {Casse}}]{Panoglou+12}
{Panoglou}, D., {Cabrit}, S., {Pineau Des For{\^e}ts}, G., {et~al.} 2012, \aap,
  538, A2

\bibitem[{{Pinte} {et~al.}(2016){Pinte}, {Dent}, {M{\'e}nard}, {Hales}, {Hill},
  {Cortes}, \& {de Gregorio-Monsalvo}}]{Pinte+16}
{Pinte}, C., {Dent}, W.~R.~F., {M{\'e}nard}, F., {et~al.} 2016, \apj, 816, 25

\bibitem[{{Pollack} {et~al.}(1985){Pollack}, {McKay}, \&
  {Christofferson}}]{Pollack+85}
{Pollack}, J.~B., {McKay}, C.~P., \& {Christofferson}, B.~M. 1985, \icarus, 64,
  471

\bibitem[{{Press} {et~al.}(1996){Press}, {Teukolsky}, {Vetterling}, \&
  {Flannery}}]{Press+96}
{Press}, W.~H., {Teukolsky}, S.~A., {Vetterling}, W.~T., \& {Flannery}, B.~P.
  1996, {Numerical recipes in FORTRAN 90}, 2nd edn. (Cambridge: Cambridge Univ.
  Press)

\bibitem[{{Quanz} {et~al.}(2006){Quanz}, {Henning}, {Bouwman}, {Ratzka}, \&
  {Leinert}}]{Quanz+06}
{Quanz}, S.~P., {Henning}, T., {Bouwman}, J., {Ratzka}, T., \& {Leinert}, C.
  2006, \apj, 648, 472

\bibitem[{{Safier}(1993)}]{Safier93}
{Safier}, P.~N. 1993, \apj, 408, 115

\bibitem[{{Salmeron} \& {Ireland}(2012)}]{SalmeronIreland12}
{Salmeron}, R., \& {Ireland}, T.~R. 2012, Earth and Planetary Science Letters,
  327, 61

\bibitem[{{Sargent} {et~al.}(2009){Sargent}, {Forrest}, {Tayrien}, {McClure},
  {Watson}, {Sloan}, {Li}, {Manoj}, {Bohac}, {Furlan}, {Kim}, \&
  {Green}}]{Sargent+09}
{Sargent}, B.~A., {Forrest}, W.~J., {Tayrien}, C., {et~al.} 2009, \apjs, 182,
  477

\bibitem[{{Schegerer} {et~al.}(2008){Schegerer}, {Wolf}, {Ratzka}, \&
  {Leinert}}]{Schegerer+08}
{Schegerer}, A.~A., {Wolf}, S., {Ratzka}, T., \& {Leinert}, C. 2008, \aap, 478,
  779

\bibitem[{{Shu} {et~al.}(2008){Shu}, {Lizano}, {Galli}, {Cai}, \&
  {Mohanty}}]{Shu+08}
{Shu}, F.~H., {Lizano}, S., {Galli}, D., {Cai}, M.~J., \& {Mohanty}, S. 2008,
  \apjl, 682, L121

\bibitem[{{Shu} {et~al.}(2000){Shu}, {Najita}, {Shang}, \& {Li}}]{Shu+00}
{Shu}, F.~H., {Najita}, J.~R., {Shang}, H., \& {Li}, Z.-Y. 2000, in Protostars
  and Planets IV, ed. V.~Mannings, A.~Boss, \& S.~Russell (Tucson, AZ: Univ.
  Arizona Press), 789

\bibitem[{{Shu} {et~al.}(2001){Shu}, {Shang}, {Gounelle}, {Glassgold}, \&
  {Lee}}]{Shu+01}
{Shu}, F.~H., {Shang}, H., {Gounelle}, M., {Glassgold}, A.~E., \& {Lee}, T.
  2001, \apj, 548, 1029

\bibitem[{{Shu} {et~al.}(1996){Shu}, {Shang}, \& {Lee}}]{Shu+96}
{Shu}, F.~H., {Shang}, H., \& {Lee}, T. 1996, Science, 271, 1545

\bibitem[{{Sicilia-Aguilar} {et~al.}(2009){Sicilia-Aguilar}, {Bouwman},
  {Juh{\'a}sz}, {Henning}, {Roccatagliata}, {Lawson}, {Acke}, {Feigelson},
  {Tielens}, {Decin}, \& {Meeus}}]{Sicilia-Aguilar+09}
{Sicilia-Aguilar}, A., {Bouwman}, J., {Juh{\'a}sz}, A., {et~al.} 2009, \apj,
  701, 1188

\bibitem[{{Simon} {et~al.}(2018){Simon}, {Bai}, {Flaherty}, \&
  {Hughes}}]{Simon+18}
{Simon}, J.~B., {Bai}, X.-N., {Flaherty}, K.~M., \& {Hughes}, A.~M. 2018, \apj,
  865, 10

\bibitem[{{Takeuchi} \& {Lin}(2002)}]{TakeuchiLin02}
{Takeuchi}, T., \& {Lin}, D.~N.~C. 2002, \apj, 581, 1344

\bibitem[{{Teitler}(2011)}]{Teitler11}
{Teitler}, S. 2011, \apj, 733, 57

\bibitem[{{Tilling} {et~al.}(2012){Tilling}, {Woitke}, {Meeus}, {Mora},
  {Montesinos}, {Riviere-Marichalar}, {Eiroa}, {Thi}, {Isella}, {Roberge},
  {Martin-Zaidi}, {Kamp}, {Pinte}, {Sandell}, {Vacca}, {M{\'e}nard},
  {Mendigut{\'{\i}}a}, {Duch{\^e}ne}, {Dent}, {Aresu}, {Meijerink}, \&
  {Spaans}}]{Tilling+12}
{Tilling}, I., {Woitke}, P., {Meeus}, G., {et~al.} 2012, \aap, 538, A20

\bibitem[{{Turner} {et~al.}(2014){Turner}, {Fromang}, {Gammie}, {Klahr},
  {Lesur}, {Wardle}, \& {Bai}}]{Turner+14}
{Turner}, N.~J., {Fromang}, S., {Gammie}, C., {et~al.} 2014, in Protostars and
  Planets VI, ed. {H. Beuther et al.} (Tucson, AZ: Univ. Arizona Press), 411

\bibitem[{{van Boekel} {et~al.}(2005){van Boekel}, {Min}, {Waters}, {de Koter},
  {Dominik}, {van den Ancker}, \& {Bouwman}}]{vanBoekel+05}
{van Boekel}, R., {Min}, M., {Waters}, L.~B.~F.~M., {et~al.} 2005, \aap, 437,
  189

\bibitem[{{van Boekel} {et~al.}(2004){van Boekel}, {Min}, {Leinert}, {Waters},
  {Richichi}, {Chesneau}, {Dominik}, {Jaffe}, {Dutrey}, {Graser}, {Henning},
  {de Jong}, {K{\"o}hler}, {de Koter}, {Lopez}, {Malbet}, {Morel}, {Paresce},
  {Perrin}, {Preibisch}, {Przygodda}, {Sch{\"o}ller}, \&
  {Wittkowski}}]{vanBoekel+04}
{van Boekel}, R., {Min}, M., {Leinert}, C., {et~al.} 2004, \nat, 432, 479

\bibitem[{{Vinkovi{\'c}}(2009)}]{Vinkovic09}
{Vinkovi{\'c}}, D. 2009, \nat, 459, 227

\bibitem[{{Wardle} \& {K{\"o}nigl}(1993)}]{WardleKonigl93}
{Wardle}, M., \& {K{\"o}nigl}, A. 1993, \apj, 410, 218

\bibitem[{{Watson} {et~al.}(2009){Watson}, {Leisenring}, {Furlan}, {Bohac},
  {Sargent}, {Forrest}, {Calvet}, {Hartmann}, {Nordhaus}, {Green}, {Kim},
  {Sloan}, {Chen}, {Keller}, {d'Alessio}, {Najita}, {Uchida}, \&
  {Houck}}]{watson+09}
{Watson}, D.~M., {Leisenring}, J.~M., {Furlan}, E., {et~al.} 2009, \apjs, 180,
  84

\bibitem[{{Weingartner} \& {Draine}(2001{\natexlab{a}})}]{WeingartnerDraine01a}
{Weingartner}, J.~C., \& {Draine}, B.~T. 2001{\natexlab{a}}, \apj, 548, 296

\bibitem[{{Weingartner} \& {Draine}(2001{\natexlab{b}})}]{WeingartnerDraine01b}
---. 2001{\natexlab{b}}, \apjs, 134, 263

\bibitem[{{Wong} {et~al.}(2016){Wong}, {Hirashita}, \& {Li}}]{Wong+16}
{Wong}, Y.~H.~V., {Hirashita}, H., \& {Li}, Z.-Y. 2016, \pasj, 68, 67

\bibitem[{{Youdin} \& {Lithwick}(2007)}]{YoudinLithwick07}
{Youdin}, A.~N., \& {Lithwick}, Y. 2007, \icarus, 192, 588

\bibitem[{{Zhu} {et~al.}(2007){Zhu}, {Hartmann}, {Calvet}, {Hernandez},
  {Muzerolle}, \& {Tannirkulam}}]{Zhu+07}
{Zhu}, Z., {Hartmann}, L., {Calvet}, N., {et~al.} 2007, \apj, 669, 483

\bibitem[{{Zolensky} {et~al.}(2006){Zolensky}, {Zega}, {Yano}, {Wirick},
  {Westphal}, {Weisberg}, {Weber}, {Warren}, {Velbel}, {Tsuchiyama}, {Tsou},
  {Toppani}, {Tomioka}, {Tomeoka}, {Teslich}, {Taheri}, {Susini}, {Stroud},
  {Stephan}, {Stadermann}, {Snead}, {Simon}, {Simionovici}, {See}, {Robert},
  {Rietmeijer}, {Rao}, {Perronnet}, {Papanastassiou}, {Okudaira}, {Ohsumi},
  {Ohnishi}, {Nakamura-Messenger}, {Nakamura}, {Mostefaoui}, {Mikouchi},
  {Meibom}, {Matrajt}, {Marcus}, {Leroux}, {Lemelle}, {Le}, {Lanzirotti},
  {Langenhorst}, {Krot}, {Keller}, {Kearsley}, {Joswiak}, {Jacob}, {Ishii},
  {Harvey}, {Hagiya}, {Grossman}, {Grossman}, {Graham}, {Gounelle}, {Gillet},
  {Genge}, {Flynn}, {Ferroir}, {Fallon}, {Ebel}, {Dai}, {Cordier}, {Clark},
  {Chi}, {Butterworth}, {Brownlee}, {Bridges}, {Brennan}, {Brearley},
  {Bradley}, {Bleuet}, {Bland}, \& {Bastien}}]{Zolensky+06}
{Zolensky}, M.~E., {Zega}, T.~J., {Yano}, H., {et~al.} 2006, Science, 314, 1735

\bibitem[{{Zsom} {et~al.}(2011){Zsom}, {Ormel}, {Dullemond}, \&
  {Henning}}]{Zsom+11}
{Zsom}, A., {Ormel}, C.~W., {Dullemond}, C.~P., \& {Henning}, T. 2011, \aap,
  534, A73

\end{thebibliography}
\end{document}